\documentclass[sigconf, screen]{acmart}

\setcopyright{none}
\acmConference[ISCA '27]{54th IEEE/ACM International Symposium on Computer Architecture}{June 5--9, 2027}{Atlanta, Georgia, USA}
\acmYear{2027}
\acmISBN{978-x-xxxx-xxxx-x/YY/MM}
\acmDOI{XXXXXXX.XXXXXXX}
\usepackage{array}
\usepackage{multirow}
\usepackage{textcomp}
\usepackage{tikz,pgfplots}
\usepackage{enumitem}
\usepackage{stfloats}
\pgfplotsset{compat=1.17}
\usetikzlibrary{positioning,calc,decorations.pathreplacing,arrows.meta,shapes.misc,patterns,patterns.meta}
\usepgfplotslibrary{groupplots}
\definecolor{lpBlue}{HTML}{0072B2}
\definecolor{lpVerm}{HTML}{D55E00}
\definecolor{lpGrey}{HTML}{8A8F98}
\definecolor{lpVic}{HTML}{E69F00}
\definecolor{lpRed}{HTML}{C0392B}
\newcommand{\lpTick}{\fontsize{7}{8}\selectfont\sffamily}
\newcommand{\lpLab}{\fontsize{8}{9}\selectfont\sffamily}
\newcommand{\lpLeg}{\fontsize{6.5}{7.5}\selectfont\sffamily}
\tikzset{lpOrd/.style={black!45, dashed, line width=0.5pt},
  lpdash/.style={dash pattern=on 3pt off 2pt},
  lpHatch/.style={postaction={pattern={Lines[angle=45, distance=2.8pt, line width=0.45pt]}, pattern color=lpVerm!75!black}}}
\pgfplotsset{
  lpbase/.style={font=\lpTick, label style={font=\lpLab}, tick label style={font=\lpTick},
    title style={font=\lpLab\bfseries, yshift=-1.5mm},
    grid=major, grid style={black!8}, axis line style={black!50}, tick style={black!50},
    every axis plot/.append style={line width=0.9pt}, mark size=2.2pt,
    legend style={font=\lpLeg, draw=none, fill=none, /tikz/every even column/.append style={column sep=2.5mm}},
    legend cell align=left},
  lpCtlS/.style={black, only marks, mark=*, mark options={solid, fill=black, line width=0.8pt}},
  lpCtlV/.style={black, only marks, mark=o, mark options={solid, fill=white, line width=0.9pt}},
  lpRawS/.style={lpVerm, only marks, mark=square*, mark options={solid, fill=lpVerm, line width=0.8pt}},
  lpRawV/.style={lpVerm, only marks, mark=square, mark options={solid, fill=white, line width=0.9pt}},
  lpOra/.style={lpBlue, only marks, mark=-, mark size=3.2pt, mark options={line width=1.5pt}},
  lpSrS/.style={lpGrey, only marks, mark=diamond*, mark size=2.6pt, mark options={solid, fill=lpGrey, line width=0.8pt}},
  lpSrV/.style={lpGrey, only marks, mark=diamond, mark size=2.6pt, mark options={solid, fill=white, line width=0.9pt}},
  lpVicS/.style={lpVic, only marks, mark=triangle*, mark size=2.6pt, mark options={solid, fill=lpVic, line width=0.8pt}},
  lpVicV/.style={lpVic, only marks, mark=triangle, mark size=2.6pt, mark options={solid, fill=white, line width=0.9pt}},
  lpLeafLS/.style={lpBlue, solid, line width=0.9pt, mark=*, mark options={solid, fill=lpBlue}},
  lpLeafLV/.style={lpBlue!55, solid, line width=0.9pt, mark=o, mark options={solid, fill=white}},
  lpRawLS/.style={lpVerm, lpdash, line width=0.9pt, mark=square*, mark options={solid, fill=lpVerm}},
  lpRawLV/.style={lpVerm!55, lpdash, line width=0.9pt, mark=square, mark options={solid, fill=white}},
  lpCtlLS/.style={black, solid, line width=1.0pt, mark=*, mark options={solid, fill=black}},
  lpCtlLV/.style={black!50, solid, line width=1.0pt, mark=o, mark options={solid, fill=white}},
}

\pgfplotsset{table/search path={.,figures/data}}

\begin{document}

\title{Pack Light, Walk Fast: Compact Page-Table Residency in Shared Caches}

\author{Samarth Narang\textsuperscript{1,2}\quad Aamer Jaleel\textsuperscript{2}\quad Brandon Reagen\textsuperscript{1}}
\affiliation{\institution{\textsuperscript{1}New York University\qquad \textsuperscript{2}NVIDIA}\country{}}

\renewcommand{\shortauthors}{Narang et al.}

\begin{abstract}
Hardware page-table walkers fetch descriptors through the cache hierarchy, causing page-table walks to compete directly with application data for shared-cache capacity. Retaining leaf page-table lines on chip avoids costly off-chip DRAM reads during address translation, yet as a workload's mapped footprint grows, those retained leaf lines displace application data and slow down cache-sensitive co-runners. We introduce LeafPack, a shared-cache organization that resolves this trade-off by compressing architectural page-table leaf lines in place within standard cache ways. By exploiting regularity in the physical frame numbers and attributes of neighboring descriptors, we pack up to sixteen 64-byte leaf lines into a single 80-byte slot inclusive of metadata. Our cache layout packs incoming leaf lines incrementally as walks fetch them from memory and reconstructs individual 64-byte lines on demand while leaving irregular lines uncompressed. This organization preserves both the in-memory page-table format and the hardware page-table walk algorithm, requiring only that walkers identify their requests as leaf accesses. To adapt page-table occupancy to changing workload demands, we pair this layout with a runtime set-dueling controller that arbitrates between compressed and uncompressed residency ceilings to balance translation coverage against application-data capacity. We evaluate LeafPack in full-system gem5 simulations of an Arm platform across both a small core and a server-class out-of-order core on workloads spanning one to eight cores. Across nine translation-heavy workloads, we improve geometric-mean harmonic speedup over an ordinary shared cache by 28.7\% on the small core and 28.4\% on the server-class core. Against an ordinary cache using SRRIP replacement, we improve geometric-mean harmonic speedup on these nine workloads by 18.7\% and 17.7\%. On twenty unseen held-out mixes evaluated after freezing the controller, LeafPack improves harmonic speedup over an ordinary cache on every mix, achieving geometric-mean gains of 21.4\% on the small core and 18.5\% on the server-class core.
\end{abstract}

\maketitle

\section{Introduction}


When a memory access misses in the TLB hierarchy, the hardware page-table walker resolves the translation through dependent descriptor accesses~\cite{bhattacharjee2017virtualmemory,bhattacharjee2017translationwall}. Page-walk caches can satisfy or bypass upper-level accesses, shortening the walk, but accesses to leaf descriptors can still miss in the on-chip cache hierarchy and stall the walk on main memory. While upper-level descriptors cover wide spans of virtual address space with relatively few entries, the leaf level grows linearly with the mapped footprint, as mapping a dense 4~GiB region with standard 4~KiB pages requires more than one million eight-byte leaf descriptors totaling 8~MiB. When repeatedly accessed leaf lines outgrow the shared cache, walkers must fetch evicted lines from DRAM even if the upper levels remain on chip. Retaining those leaf lines in the cache avoids off-chip walk accesses, yet every line dedicated to page-table metadata displaces application data that cores also need.

Architects have attacked translation overhead from both the TLB side and the cache side. Enlarging TLBs, coalescing contiguous translations as in CoLT~\cite{colt}, and adding concurrent hardware walkers reduce walk frequency or overlap independent misses. The walks that still remain must fetch descriptors through the memory hierarchy, where cache replacement policies such as EWaH and xPTP~\cite{ewah,xptp} reduce walk latency by preferentially retaining page-table lines in the data caches. Because these policies store leaf lines uncompressed, protecting a growing leaf footprint starves co-running applications of data capacity, and any leaf footprint that exceeds the cache size cannot fit without compression.

We observe that although page-table leaf lines are reused, the cache space occupied by them holds redundant information. Every eight-byte leaf descriptor combines a physical frame number with permissions and status bits, and neighboring descriptors frequently share identical attributes while referencing contiguous physical frames. Whenever the eight descriptors inside a 64-byte leaf line follow a regular progression, the line's first descriptor and stride determine the remaining seven entries, and neighboring leaf lines that share a common base frame allow their starting descriptors to be represented by small offsets. We exploit this spatial regularity inside the shared cache through compression to both accelerate standalone execution by retaining more active leaves than uncompressed protection allows and preserve application-data capacity by holding the same leaves in fewer ways.

We build on this insight to design \emph{LeafPack}, a shared-cache organization that packs up to sixteen 64-byte leaf lines into an 80-byte slot inclusive of metadata. As hardware walkers fetch leaf lines from memory, LeafPack incrementally encodes regular lines into shared slots and reconstructs individual lines on demand while leaving irregular lines uncompressed. Pairing four payload ways with a fifth metadata way preserves the physical cache array, where per-line coherence bits maintain independent invalidation across sibling lines. We only require hardware walkers to tag their page-table and leaf requests, leaving the page-table format, walk algorithm, and operating-system TLB shootdowns unchanged.

LeafPack packs leaf lines incrementally as misses arrive while preserving per-line coherence. Because even a partially filled group occupies five physical ways, we charge all converted ways, including metadata and empty slots, when balancing translation coverage against application-data capacity.

\definecolor{lvA}{HTML}{3B4B8C}   
\definecolor{lvB}{HTML}{3F7CC4}   
\definecolor{lvC}{HTML}{2A9D8F}   
\definecolor{lvD}{HTML}{E08A1E}   
\definecolor{capc}{HTML}{C0392B}  
\definecolor{fitc}{HTML}{2A9D8F}  
\tikzset{
  ent/.style={draw=#1!80!black, fill=#1!12, minimum width=5mm, minimum height=4.4mm,
              inner sep=0pt, font=\scriptsize\sffamily, line width=0.4pt},
  sel/.style={draw=#1!80!black, fill=#1!75, text=white, minimum width=5mm, minimum height=4.4mm,
              inner sep=0pt, font=\scriptsize\sffamily\bfseries, line width=0.6pt},
  dots/.style={minimum width=4mm, minimum height=4.4mm, inner sep=0pt, font=\scriptsize},
  ptr/.style={-{Latex[length=1.5mm]}, line width=0.6pt, draw=#1!80!black},
}
\begin{figure}[t]
\centering
\begin{tikzpicture}
\begin{axis}[
  width=6.7cm, height=4.0cm,
  ybar=0pt, bar width=6pt,
  ymode=log, log origin=infty, log basis y=2, ymin=2e-3, ymax=6000,
  ytick={0.0078125,0.125,2,32,512},
  yticklabels={8\,KiB,128\,KiB,2\,MiB,32\,MiB,512\,MiB},
  xmin=0.5, xmax=6.5, xtick={1,...,6},
  xticklabels={256\,MiB,1\,GiB,4\,GiB,16\,GiB,64\,GiB,256\,GiB},
  tick label style={font=\scriptsize\sffamily},
  label style={font=\scriptsize\sffamily},
  xlabel={mapped footprint (4\,KiB pages)}, xlabel style={yshift=1mm},
  ylabel={page-table bytes}, ylabel style={yshift=-2mm},
  axis line style={gray!60}, tick style={gray!60}, major tick length=2pt,
  legend style={font=\scriptsize\sffamily, draw=none, fill=white, fill opacity=0.9, text opacity=1,
                at={(0.02,0.98)}, anchor=north west, cells={anchor=west},
                legend image code/.code={\fill[#1] (0,-0.6mm) rectangle (2.4mm,1.2mm);}},
  clip=false,
]
\fill[fitc!14] (axis cs:0.5,0.002) rectangle (axis cs:6.5,0.5);
\fill[fitc!6]  (axis cs:0.5,0.5) rectangle (axis cs:6.5,4);
\addplot[fill=lvC!60!black, draw=lvC!50!black] coordinates
  {(1,0.01172) (2,0.01172) (3,0.02344) (4,0.07031) (5,0.25781) (6,1.0078)};
\addplot[fill=lvD, draw=lvD!80!black,
  nodes near coords, point meta=explicit symbolic,
  every node near coord/.append style={font=\scriptsize\sffamily\bfseries, text=capc, fill=white, inner sep=1pt, yshift=1pt}]
  coordinates {(1,0.5) [$\times$1] (2,2) [$\times$4] (3,8) [$\times$16] (4,32) [$\times$64]
               (5,128) [$\times$256] (6,512) [$\times$1024]};
\legend{L0 to L2 tables, L3 leaves}
\draw[capc, line width=1pt, dashed] (axis cs:0.5,0.5) -- (axis cs:6.5,0.5);
\draw[capc, line width=1pt] (axis cs:0.5,4) -- (axis cs:6.5,4);
\node[font=\scriptsize\sffamily\bfseries, text=capc, anchor=north west, align=left, inner sep=1pt] at (axis cs:6.55,0.5) {0.5\,MiB\\per core};
\node[font=\scriptsize\sffamily\bfseries, text=capc, anchor=south west, align=left, inner sep=1pt] at (axis cs:6.55,4) {4\,MiB\\8 cores};
\end{axis}
\end{tikzpicture}
\caption{\textbf{Page-table storage versus mapped footprint.}}
\label{fig:motivation}
\end{figure}
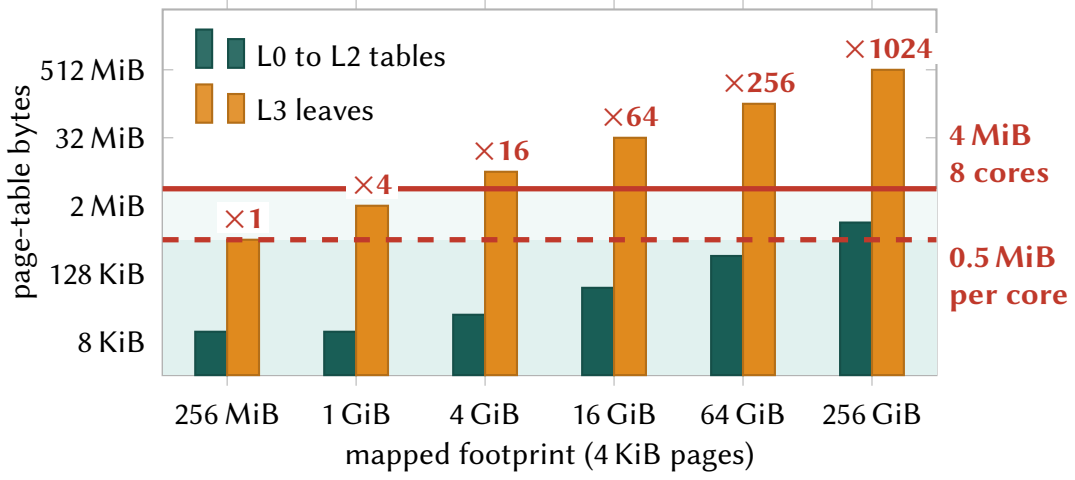

LeafPack uses set dueling to choose between two compressed way ceilings and one uncompressed ceiling as translation and data demands shift. Physical leader sets evaluate the compressed policies while shadow tags evaluate the uncompressed policy, and per-core miss weighting prevents high-traffic cores from dominating selection. Although one could combine prior page-table compression~\cite{panwar2022tmcc}, shared-tag layouts~\cite{sardashti2016yacc}, and adaptive cache partitioning~\cite{qureshi2006ucp,csalt}, compressing each page-table line on its own inside a shared-tag layout under our controller barely improves over uncompressed protection (Section~\ref{sec:eval-breakdown}). LeafPack instead shares a single base descriptor across neighboring leaf lines, packing far more page-table coverage into a bounded physical allocation.

We evaluate LeafPack in full-system Arm gem5 simulations on small and server cores across one to eight cores. Our baseline shared L2 stores page-table lines and application data without compression or page-table protection, using random replacement unless noted. On nine translation-heavy workloads, LeafPack improves geometric-mean harmonic speedup over the SRRIP baseline by 18.7\% and 17.7\% on the small and server cores. Across all 18 development workloads under random replacement, it gains 4.7\% and 4.8\% over the best static uncompressed configuration (including baseline fallback), or about 2.0\% and 2.1\% when we exclude standalone 4~GiB SysBench. Across twenty held-out mixes, our frozen controller improves every mix over the baseline with geometric-mean gains of 21.4\% and 18.5\%, and beats the best static uncompressed configuration by 2.9\% and 2.1\%.

This paper makes three contributions:
\begin{itemize}[topsep=0pt]
  \item \textbf{Compressed leaf residency.} We design a shared-cache organization that packs neighboring architectural leaf lines into five-way groups using a shared base descriptor while preserving per-line coherence and falling back to uncompressed storage when needed.
  \item \textbf{Adaptive way selection.} We design an online controller that extends set dueling with shadow tags and per-core miss weighting to select between compressed and uncompressed page-table way ceilings.
  \item \textbf{Full-system evaluation.} We evaluate LeafPack across single-program and multiprogrammed workloads to separate the gains of protection, compression, and online selection, compare against prior translation-caching designs, and examine held-out mixes and controller failure cases.
\end{itemize}

\section{Background and Motivation}\label{sec:background}

With 4~KiB pages and 8-byte descriptors, each 64-byte leaf line maps 32~KiB of virtual memory. While upper-level entries cover larger address ranges and fit inside small walk caches, reused leaf lines can outgrow the shared cache and force off-chip DRAM accesses (Fig.~\ref{fig:motivation}).

\textit{Translation competes with data.}
Prior architectures use the on-chip cache hierarchy in several ways to cut down the cost of page-table walks. Page-table-aware replacement policies such as EWaH and xPTP~\cite{ewah,xptp} identify cache lines fetched by hardware walkers and retain them longer than application-data lines, while Victima~\cite{victima} repurposes L2 cache blocks to hold clusters of completed translations evicted from the TLBs. Whether a mechanism retains raw page-table lines or synthesized translations, it consumes physical cache ways that would otherwise hold application data. Its net benefit therefore depends on whether the off-chip walk accesses it avoids outweigh the additional application-data misses caused by shrinking the effective data cache, especially when a translation-heavy workload shares the cache with a data-sensitive co-runner.

\textit{Regularity creates a capacity opportunity.}
The internal structure of page-table leaf lines offers a natural way to break this capacity trade-off through compression. Because neighboring eight-byte leaf descriptors regularly share permission and status flags while referencing contiguous or nearby physical frames, multiple 64-byte leaf lines can be represented compactly using shared base descriptors and small frame offsets (Fig.~\ref{fig:compression}). Turning this regularity into real shared-cache capacity savings requires packing leaf lines incrementally as walkers fetch them on demand, maintaining per-line coherence, falling back to uncompressed storage for irregular mappings, and returning emptied ways to application data.

\definecolor{lvD}{HTML}{E08A1E}   
\definecolor{pk}{HTML}{2A9D8F}    
\definecolor{dn}{HTML}{6A4C93}    
\definecolor{bad}{HTML}{C0392B}   
\definecolor{mt}{HTML}{8A8F98}    
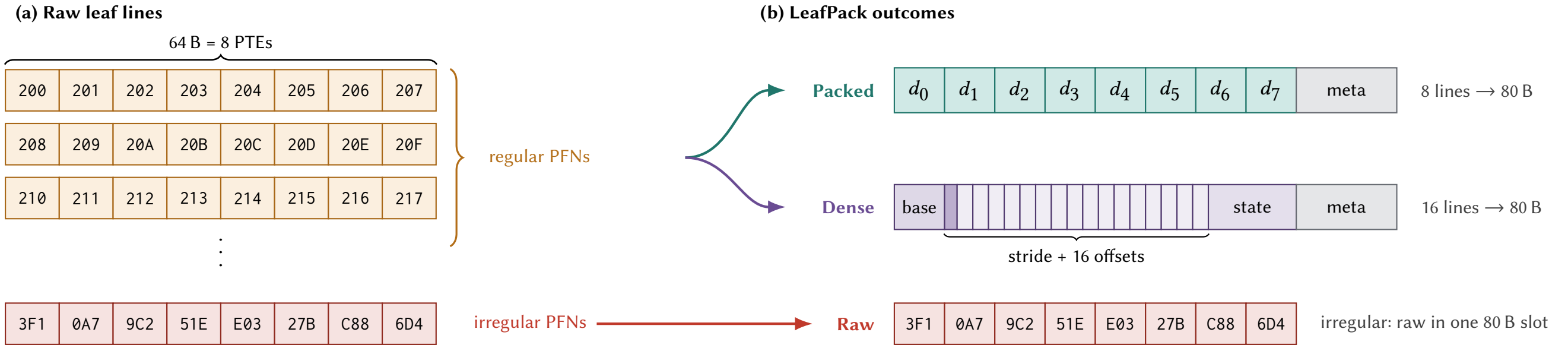
\begin{figure*}[t]
\centering
\begin{tikzpicture}[
  font=\scriptsize\sffamily,
  pte/.style={draw=#1!75!black, fill=#1!14, minimum width=6mm, minimum height=4.6mm, inner sep=0pt,
              font=\scriptsize\ttfamily, line width=0.4pt},
  fld/.style={draw=#1!75!black, fill=#1!22, minimum height=5mm, inner sep=0pt, line width=0.4pt,
              font=\scriptsize\sffamily, anchor=west},
  hd/.style={font=\scriptsize\sffamily\bfseries, anchor=west},
  nm/.style={font=\scriptsize\sffamily\bfseries, anchor=east, text=#1},
  note/.style={font=\scriptsize\sffamily, anchor=west, text=black!75},
  flow/.style={-{Latex[length=2mm]}, line width=0.8pt, draw=#1},
]
\node[hd] at (-0.3,0.85) {(a) Raw leaf lines};
\newcommand{\leafrow}[4]{
  \foreach \p [count=\c from 0] in {#3} {\node[pte=#2] (#4\c) at ({0.6*\c},#1) {\p};}}
\leafrow{0}{lvD}{200,201,202,203,204,205,206,207}{ra}
\leafrow{-0.6}{lvD}{208,209,20A,20B,20C,20D,20E,20F}{rb}
\leafrow{-1.2}{lvD}{210,211,212,213,214,215,216,217}{rc}
\node at (2.1,-1.7) {$\vdots$};
\leafrow{-2.6}{bad}{3F1,0A7,9C2,51E,E03,27B,C88,6D4}{rx}
\draw[decorate, decoration={brace, amplitude=3pt}, line width=0.5pt]
  ([yshift=0.5mm]ra0.north west) -- ([yshift=0.5mm]ra7.north east)
  node[midway, above=2pt] {64\,B = 8 PTEs};
\draw[decorate, decoration={brace, amplitude=4pt}, line width=0.6pt, draw=lvD!80!black]
  ([xshift=1.5mm]ra7.north east) -- ([xshift=1.5mm, yshift=-3mm]rc7.south east)
  coordinate[midway, xshift=2.2mm] (reg);
\node[font=\scriptsize\sffamily, text=lvD!80!black, anchor=west] at ([xshift=1mm]reg) {regular PFNs};
\node[font=\scriptsize\sffamily, text=bad, anchor=west] (irr) at ([xshift=3mm]rx7.east) {irregular PFNs};

\begin{scope}[xshift=9.6cm]
\node[hd] at (-1.6,0.85) {(b) LeafPack outcomes};
\node[nm=pk!75!black] at (-0.1,0) {Packed};
\foreach \i in {0,...,7} {\node[fld=pk, minimum width=5.6mm] (pd\i) at ({0.56*\i},0) {$d_{\textstyle\i}$};}
\node[fld=mt, minimum width=11.2mm] (pm) at (4.48,0) {meta};
\node[note] at ([xshift=1.5mm]pm.east) {8 lines $\to$ 80\,B};
\node[nm=dn] at (-0.1,-1.3) {Dense};
\node[fld=dn, minimum width=5.6mm] (db) at (0,-1.3) {base};
\node[fld=dn, fill=dn!45, minimum width=1.4mm] (ds) at (0.56,-1.3) {};
\foreach \i in {0,...,15} {\node[fld=dn, fill=dn!10, minimum width=1.75mm] (dd\i) at ({0.70+0.175*\i},-1.3) {};}
\node[fld=dn, fill=dn!18, minimum width=9.8mm] (dst) at (3.5,-1.3) {state};
\node[fld=mt, minimum width=11.2mm] (dm) at (4.48,-1.3) {meta};
\draw[decorate, decoration={brace, amplitude=2pt, mirror}, line width=0.4pt]
  ([yshift=-0.4mm]ds.south west) -- ([yshift=-0.4mm]dd15.south east)
  node[midway, below=1.5pt, font=\scriptsize\sffamily] {stride + 16 offsets};
\node[note] at ([xshift=1.5mm]dm.east) {16 lines $\to$ 80\,B};
\node[nm=bad] at (-0.1,-2.6) {Raw};
\foreach \p [count=\c from 0] in {3F1,0A7,9C2,51E,E03,27B,C88,6D4}
  {\node[pte=bad, minimum width=5.6mm, anchor=west] (fx\c) at ({0.56*\c},-2.6) {\p};}
\node[note] at ([xshift=1.5mm]fx7.east) {irregular: raw in one 80\,B slot};
\end{scope}

\draw[flow=pk!75!black] ([xshift=24mm]reg) to[out=0, in=180] ([xshift=-12mm]pd0.west);
\draw[flow=dn]          ([xshift=24mm]reg) to[out=0, in=180] ([xshift=-12mm]db.west);
\draw[flow=bad]         (irr.east) -- ([xshift=-9mm]fx0.west);
\end{tikzpicture}
\caption{\textbf{LeafPack leaf-line formats.} (a) Regular and irregular physical frame numbers (PFNs). (b) Packed, Dense, and raw slots.}
\label{fig:compression}
\end{figure*}

\section{LeafPack}\label{sec:leafpack}

A \emph{leaf line} is a 64-byte cache line containing eight 8-byte leaf page-table entries (PTEs)~\cite{arm_ddi0487,linux_pagetables}. LeafPack compresses and combines these lines to increase the shared L2's effective page-table capacity.

\subsection{Leaf Compression}

Both LeafPack formats (Fig.~\ref{fig:compression}) exploit regularity within a leaf line. When all eight PTEs are valid page descriptors that map consecutive frames $f, f+1, \dots, f+7$ and match in every bit outside the frame field (bits $[63:48]$ and $[11:0]$, including the access flag, dirty state, and software-defined bits), storing the first 8-byte PTE verbatim determines the remaining seven and reduces the 64-byte leaf line to an 8-byte descriptor. Any other line, including one with an invalid PTE, stays uncompressed.

\emph{Packed format.}
Packed stores the first descriptors of up to eight leaf lines in a 64-byte payload. With 16 bytes of metadata, this payload forms an 80-byte logical \emph{slot}, representing 512 bytes at $6.4\times$ compression when full. Each line retains its own starting frame and attributes, so Packed requires no contiguity across line boundaries.

\emph{Dense format.}
Dense represents up to sixteen leaf lines using one 64-bit base descriptor, a 16-bit intra-line stride, and sixteen signed 20-bit frame offsets. The base stores the first PTE of the first line placed in the slot, each offset gives a line's starting frame relative to that base ($\pm 2^{19}$ frames, or $\pm 2$~GiB), and the stride sets each line's remaining frames (fixed to one in our evaluation). A full Dense slot represents 1,024 bytes in the same 80-byte footprint for $12.8\times$ compression, and its offsets can bridge gaps between physical allocations as long as each line is internally contiguous and shares the base's non-frame bits.

\subsection{Cache Organization}

\emph{Physical storage.}
LeafPack distributes four logical slots across a \emph{five-way group} within one cache set: four ways hold 64-byte payloads, and the fifth holds four 16-byte metadata records. This arrangement preserves the physical cache-array and bus widths.

A runtime policy selects compressed or uncompressed storage and a per-set way ceiling $W$ for the page-table class (Section~\ref{sec:controller}). The ceiling reserves no capacity: application data can use any way not held by page tables. A converted group counts as five page-table ways even when partially full, since its shared metadata way prevents application data from using empty slots. Section~\ref{sec:eval-single} describes partial groups used in static experiments.

\emph{Regions and slot tags.}
A \emph{region} is an aligned 1~KiB range of physical memory holding PTEs, distinct from the physical frames those PTEs map. A compressed slot holds lines from a single region at fixed positions indexed by address bits $[9:6]$, where Dense stores position $p$'s offset in field $p$ and Packed stores position $p$'s descriptor in word $p \bmod 8$ so that each Packed slot covers one 512-byte half of the region. Any subset of those positions can be valid at a time, and while a region may span multiple slots, each line position is valid in at most one slot.

The address bits above the 1~KiB boundary form the \emph{region prefix}, which identifies the region during lookup (Section~\ref{sec:access}). Each slot's 61-bit tag record contains this prefix and sixteen per-line valid bits. Each valid bit indicates cache residency of the corresponding line, independently of PTE validity. These records reside in the tag array, outside the 80-byte slot footprint. The slot's 16-byte metadata record holds per-line coherence state and per-slot replacement state, using 71 of its 128 bits. A three-bit \emph{group bitmap} per set identifies converted five-way groups.

\subsection{Cache Access and Replacement}\label{sec:access}

\emph{Lookup.}
An L2 lookup checks the line's normal set for uncompressed storage and its \emph{packed set} for either compressed format. The packed-set index uses only the region prefix, excluding the low ten address bits, so all sixteen line positions map to the same set. Address bits $[9:6]$ select the line position while bits $[5:0]$ select the byte.

The group bitmap restricts slot probes to converted groups. A compressed hit requires a matching region prefix and a valid bit of one for the requested line position. The cache selects that line's stored descriptor or frame offset and reconstructs the original 64-byte line, adding two cycles for Packed or three for Dense. Unconverted sets retain the standard L2 path with no added latency.

\emph{Fill and allocation.}
L2 walk misses fetch individual 64-byte lines from memory. Under a compressed policy, the first eligible leaf line in a region opens a Dense slot and sets its base descriptor. Later eligible lines from that region join the Dense slot whenever their non-frame bits match the base and their starting frame fits the 20-bit offset range. If a line cannot join the region's Dense slot, it goes to a Packed slot for that region instead, either joining one with space or opening a new one. Upper-level page-table lines and leaves with irregular or invalid PTEs stay uncompressed, taking one slot in their normal set under a compressed ceiling or a regular way under the uncompressed ceiling.

A fill needing a new slot first uses an open slot in a converted group. If none is available and the ceiling permits, the cache converts another five-way group, evicting its application data. At the ceiling, it replaces a slot using bimodal re-reference interval prediction (BRRIP)~\cite{rrip}. Application data in unconverted ways uses the baseline replacement policy.

\emph{Replacement and group release.}
Because BRRIP tracks a single re-reference value per slot rather than per line, evicting a victim slot evicts every leaf line it holds. Once evictions or invalidations clear the last valid line from a group, the cache clears its group bitmap bit and returns all five ways to application data. Lowering the way ceiling can also reclaim excess groups (Section~\ref{sec:controller}).

\subsection{Coherence and Invalidation}
\definecolor{pk}{HTML}{2A9D8F}    
\definecolor{bad}{HTML}{C0392B}   
\definecolor{mt}{HTML}{8A8F98}    
\begin{figure}[b]
\centering
\begin{tikzpicture}[
  font=\scriptsize\sffamily,
  box/.style={draw=#1!75!black, fill=#1!10, rounded corners=1.5pt, align=center,
              inner sep=2.5pt, minimum height=7mm, line width=0.5pt},
  bit/.style={draw=pk!75!black, minimum width=3.2mm, minimum height=3.2mm, inner sep=0pt,
              font=\scriptsize\sffamily, line width=0.4pt},
  line/.style={draw=pk!75!black, fill=pk!12, minimum width=8mm, minimum height=5mm, inner sep=0pt,
               line width=0.4pt},
  vb/.style={draw=pk!75!black, fill=pk!45, minimum width=3mm, minimum height=5mm, inner sep=0pt,
             font=\scriptsize\sffamily, line width=0.4pt},
  arr/.style={-{Latex[length=1.5mm]}, line width=0.6pt, draw=#1},
]
\node[box=mt, text width=19mm] (wr) at (0,0) {Page-table write\\to leaf line $X$};
\node[box=pk, minimum width=31mm, minimum height=10mm] (set) at (4.6,0) {};
\node[anchor=west] at ([xshift=1mm]set.west) {Packed set};
\node[bit, fill=white]   (b0) at ([xshift=-11.5mm, yshift=1mm]set.east) {0};
\node[bit, fill=pk!45, right=0pt of b0] (b1) {1};
\node[bit, fill=white, right=0pt of b1] (b2) {0};
\node[font=\scriptsize\sffamily, text=pk!70!black, below=0.3mm of b1, inner sep=0.5pt] {group bitmap};
\draw[arr=black!70] (wr) -- (set);

\node[box=mt, text width=19mm] (ord) at (0,-1.45) {Standard\\coherence path};
\node[box=pk, text width=27mm] (look) at (4.6,-1.45) {Group lookup\\for line $X$ in slot};
\draw[arr=mt] (set.south west) -- node[pos=0.45, above, sloped, font=\scriptsize\sffamily] {no group} (ord.north east);
\draw[arr=pk!75!black] (set) -- node[right, font=\scriptsize\sffamily] {group present} (look);

\foreach \k/\lbl/\v in {0/0/1,1/1/1,2/2/0,3/7/1} {
  \node[line] (L\k) at ({1.30+1.25*\k},-2.85) {line \lbl};
  \ifnum\v=1 \node[vb, right=0pt of L\k] (V\k) {1};
  \else \node[vb, fill=white, draw=bad, text=bad, right=0pt of L\k] (V\k) {0}; \fi}
\foreach \d in {-1.1,0,1.1} \fill[pk!70!black] ($(V2.east)!0.5!(L3.west)+(\d pt,0)$) circle (0.35pt);
\draw[bad, line width=0.9pt] ([shift={(-0.3pt,-0.3pt)}]L2.south west) rectangle ([shift={(0.3pt,0.3pt)}]V2.north east);
\coordinate (hitN) at ($(L2.north west)!0.5!(V2.north east)$);
\coordinate (hitS) at ($(L2.south west)!0.5!(V2.south east)$);
\draw[pk!75!black, dashed, line width=0.4pt, rounded corners=1.5pt]
  ([shift={(-2pt,-2pt)}]L0.south west) rectangle ([shift={(2pt,2pt)}]V3.north east);
\node[font=\scriptsize\sffamily, text=pk!70!black, anchor=south east, inner sep=1pt] at ([shift={(2pt,2pt)}]V3.north east) {compressed slot};
\node[font=\scriptsize\sffamily, text=black!60, fill=white, above=1.3mm of V0, inner sep=0.5pt] {valid};
\draw[arr=pk!75!black] (look.south) -- ([yshift=2pt]hitN);
\node[font=\scriptsize\sffamily\bfseries, text=bad, anchor=north] at ([yshift=-1.6mm]hitS) {Selective invalidation};
\end{tikzpicture}
\caption{\textbf{Coherence handling for compressed lines.}}
\label{fig:coherence}
\end{figure}
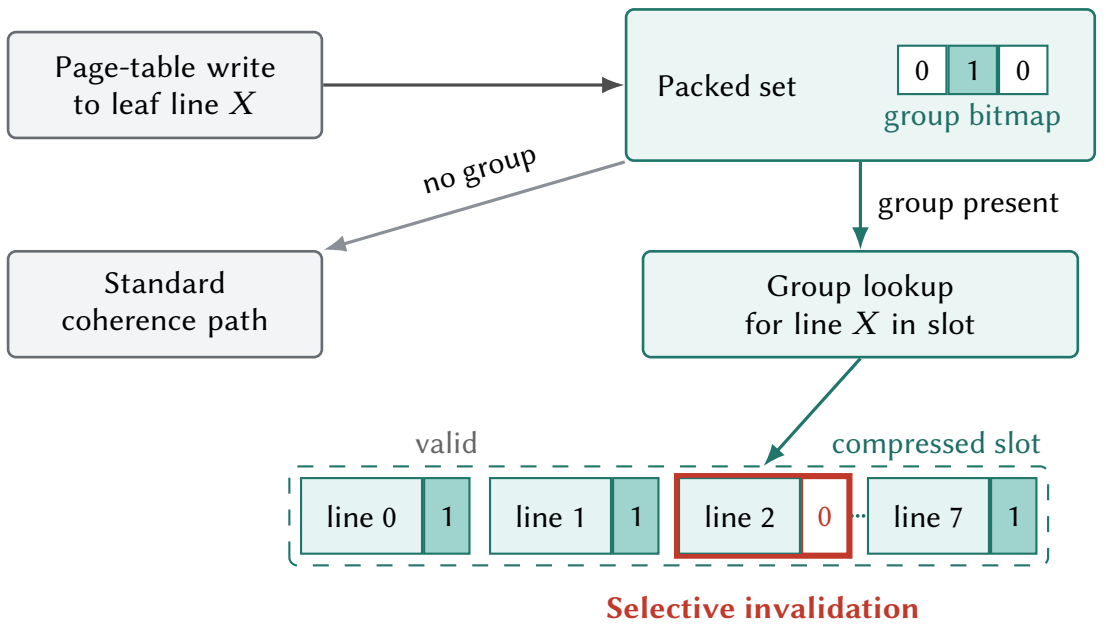

LeafPack caches architectural page-table lines without changing their layout in memory and participates in standard cache coherence~\cite{sorin2011primer}. Coherence requests locate a compressed copy using the packed set's group bitmap, region prefix, and line position (Fig.~\ref{fig:coherence}). An invalidation clears only that line's valid bit, leaving other lines in the slot accessible.

When a write to a compressed line still satisfies its slot's format, the cache updates the entry in place. Otherwise, the cache expands that line to uncompressed form inside the slot and evicts the slot's remaining lines, or writes the updated line back to memory if another line in the slot has an outstanding miss. Even when the runtime policy changes, lookups always check both the normal and packed sets so that every resident line remains reachable at a single location. Moreover, the cache never evicts a line that has an outstanding miss and executes every fill, re-encoding, or group conversion atomically on its set. Operating-system TLB shootdowns flush the TLBs as usual~\cite{miller_cachetlb} and require no LeafPack invalidation.

\subsection{Adaptive Way Selection}\label{sec:controller}
\definecolor{pk}{HTML}{2A9D8F}    
\definecolor{lvD}{HTML}{E08A1E}   
\definecolor{mt}{HTML}{8A8F98}    
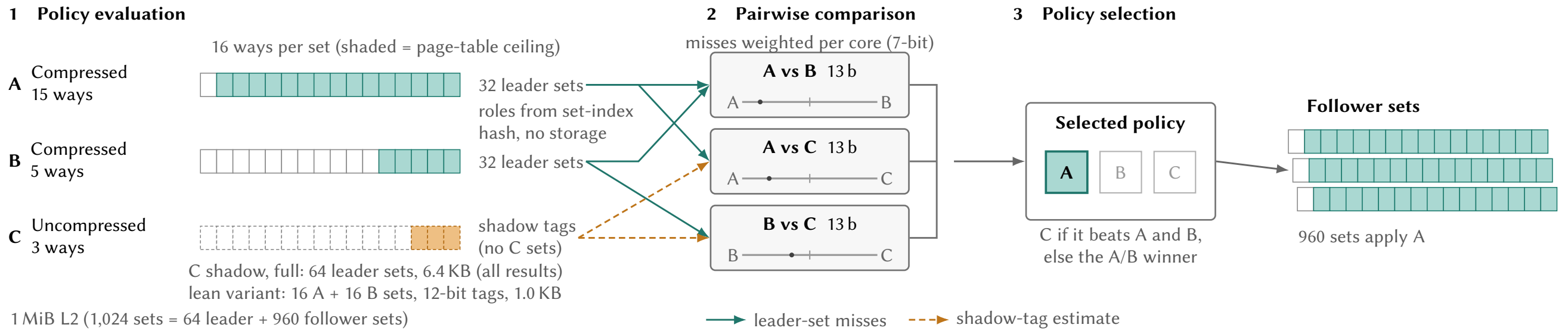
\begin{figure*}[t]
\centering
\begin{tikzpicture}[
  font=\scriptsize\sffamily,
  way/.style={draw=black!45, minimum width=1.8mm, minimum height=2.6mm, inner sep=0pt, line width=0.3pt},
  pol/.style={align=left, anchor=west, inner sep=0pt},
  ctr/.style={draw=black!55, fill=black!3, rounded corners=1.5pt, minimum width=22mm, minimum height=7.2mm,
              line width=0.5pt},
  arr/.style={-{Latex[length=1.4mm]}, line width=0.55pt, draw=#1},
  est/.style={-{Latex[length=1.4mm]}, line width=0.55pt, draw=#1, dash pattern=on 2pt off 1.2pt},
  ttl/.style={font=\scriptsize\sffamily\bfseries, anchor=west, inner sep=0pt},
  note/.style={font=\fontsize{6}{7}\selectfont\sffamily, text=black!65, inner sep=0.5pt},
]
\newcommand{\lpset}[5]{
  \foreach \q in {1,...,16} {
    \pgfmathtruncatemacro{\pt}{\q > 16-#3 ? 1 : 0}
    \ifnum\pt=1
      \ifnum#5=1 \node[way, draw=#4!80!black, dash pattern=on 1pt off 0.7pt, fill=#4!55] at ({#1+0.18*\q},#2) {};
      \else \node[way, draw=#4!80!black, fill=#4!40] at ({#1+0.18*\q},#2) {}; \fi
    \else
      \ifnum#5=1 \node[way, dash pattern=on 1pt off 0.7pt, fill=white] at ({#1+0.18*\q},#2) {};
      \else \node[way, fill=white] at ({#1+0.18*\q},#2) {}; \fi
    \fi}}

\node[ttl] at (0,1.62) {1\quad Policy evaluation};
\node[note, anchor=west] at (2.23,1.26) {16 ways per set (shaded = page-table ceiling)};
\foreach \y/\p/\d/\w/\c/\s in {0.85/A/Compressed\\15 ways/15/pk/0, 0/B/Compressed\\5 ways/5/pk/0, -0.85/C/Uncompressed\\3 ways/3/lvD/1} {
  \node[pol] at (0,\y) {\textbf{\p}\enspace\parbox[c]{17mm}{\fontsize{6}{7}\selectfont\d}};
  \lpset{2.05}{\y}{\w}{\c}{\s}
}
\node[note, anchor=west, align=left] (la) at (5.2,0.85) {32 leader sets};
\node[note, anchor=west, align=left] (lb) at (5.2,0)    {32 leader sets};
\node[note, anchor=west, align=left] (lc) at (5.2,-0.85) {shadow tags\\(no C sets)};
\node[note, anchor=west, align=left] at (5.2,0.42) {roles from set-index\\hash, no storage};
\node[note, anchor=north west, align=left] at (1.98,-1.11) {C shadow, full: 64 leader sets, 6.4\,KB (all results)\\lean variant: 16 A + 16 B sets, 12-bit tags, 1.0\,KB};
\node[note, anchor=west] at (0,-1.76) {1\,MiB L2 (1,024 sets = 64 leader + 960 follower sets)};

\node[ttl] at (7.75,1.62) {2\quad Pairwise comparison};
\foreach \y/\l/\r/\m/\n in {0.85/A/B/-0.55/ab, 0/A/C/-0.45/ac, -0.85/B/C/-0.2/bc} {
  \node[ctr] (\n) at (8.9,\y) {};
  \node[font=\scriptsize\sffamily\bfseries, anchor=north] at ([yshift=-0.6pt]\n.north) {\l\ vs \r\enspace{\mdseries 13\,b}};
  \draw[black!40, line width=0.6pt] ([xshift=-7.5mm, yshift=-1.9mm]\n.center) -- ([xshift=7.5mm, yshift=-1.9mm]\n.center);
  \draw[black!40, line width=0.4pt] ([yshift=-1.4mm]\n.center) -- ([yshift=-2.4mm]\n.center);
  \node[font=\scriptsize\sffamily, text=black!60, anchor=east, inner sep=0.8pt] at ([xshift=-7.5mm, yshift=-1.9mm]\n.center) {\l};
  \node[font=\scriptsize\sffamily, text=black!60, anchor=west, inner sep=0.8pt] at ([xshift=7.5mm, yshift=-1.9mm]\n.center) {\r};
  \fill[black!75] ($([yshift=-1.9mm]\n.center)+(\m,0)$) circle (0.75pt);
}
\node[note] at (8.9,1.31) {misses weighted per core (7-bit)};
\draw[arr=pk!75!black] (la.east) -- (ab.west);
\draw[arr=pk!75!black] (la.east) -- (7.0,0.85) -- (ac.west);
\draw[arr=pk!75!black] (lb.east) -- (7.0,0) -- (ab.west);
\draw[arr=pk!75!black] (lb.east) -- (bc.west);
\draw[est=lvD!85!black] (lc.east) -- (ac.west);
\draw[est=lvD!85!black] (lc.east) -- (bc.west);
\draw[arr=pk!75!black] (7.75,-1.76) -- ++(0.45,0) node[note, anchor=west, right=0.5mm] {leader-set misses};
\draw[est=lvD!85!black] (10.0,-1.76) -- ++(0.45,0) node[note, anchor=west, right=0.5mm] {shadow-tag estimate};

\node[ttl] at (11.15,1.62) {3\quad Policy selection};
\draw[black!55, line width=0.55pt] (ab.east) -- ++(0.3,0) |- (bc.east);
\draw[black!55, line width=0.55pt] (ac.east) -- ++(0.3,0);
\node[draw=black!55, rounded corners=1.5pt, minimum width=21mm, minimum height=13mm, line width=0.5pt] (sel) at (12.35,0) {};
\node[font=\scriptsize\sffamily\bfseries, anchor=north] at ([yshift=-1pt]sel.north) {Selected policy};
\node[draw=pk!75!black, fill=pk!40, line width=0.7pt, minimum size=4.6mm, inner sep=0pt, font=\scriptsize\sffamily\bfseries] (sa) at (11.75,-0.12) {A};
\node[draw=black!30, minimum size=4.6mm, inner sep=0pt, text=black!40] (sb) at (12.35,-0.12) {B};
\node[draw=black!30, minimum size=4.6mm, inner sep=0pt, text=black!40] (sc) at (12.95,-0.12) {C};
\draw[arr=black!60] (10.5,0) -- (sel.west);
\node[note, anchor=north, align=center] at ([yshift=-0.5mm]sel.south) {C if it beats A and B,\\else the A/B winner};

\node[font=\scriptsize\sffamily\bfseries, anchor=west] at (14.3,0.62) {Follower sets};
\foreach \k in {0,1,2} { \lpset{14.12+0.05*\k}{0.22-0.32*\k}{15}{pk}{0} }
\node[note, anchor=west] at (14.3,-0.85) {960 sets apply A};
\draw[arr=black!60] (sel.east) -- (14.22,-0.1);
\end{tikzpicture}
\caption{\textbf{Adaptive way selection.}}
\label{fig:controller}
\end{figure*}

LeafPack uses set dueling~\cite{10.1145/1250662.1250709} to select among three policies (Fig.~\ref{fig:controller}): a 15-way compressed ceiling for large active leaf sets, a 5-way compressed ceiling to protect cache-sensitive co-runners, and a 3-way uncompressed ceiling for small leaf sets.

\emph{Policy comparison.}
\emph{Leader sets} are sampled sets running candidate policies; we assign 32 physical leaders to each compressed policy. Each of these 64 leaders also has \emph{shadow tags}, tag and state entries that estimate hits under the uncompressed policy without storing data. They track up to 3 page-table entries (under SRRIP) and 16 data entries (under the baseline's data replacement policy) per set while capping the combined total at 16 valid entries to match a physical set, avoiding uncompressed leader sets that penalize translation-heavy walks.

Three saturating counters compare demand misses pairwise, weighting each miss inversely by its core's L2 access rate so high-traffic cores cannot dominate quieter co-runners. The remaining \emph{follower sets} choose the uncompressed policy when it beats both compressed policies, and the winning compressed policy otherwise.

\emph{Warm-up and probe modes.}
To let compressed leaders accumulate neighboring leaf lines, a warm-up gate freezes the two counters comparing compressed policies against shadow tags until the compressed leaders have observed $2^{15}$ accesses. Followers initially choose only between the compressed policies.

After warm-up, probe modes reduce capacity spent on losing policies. When the uncompressed policy beats both compressed policies by a quarter of the counter range, \emph{uncompressed probing} switches 30 of each compressed policy's 32 leaders to uncompressed storage, retaining two per policy to detect phase changes.

When one compressed policy beats the other by a quarter of the counter range and also beats the uncompressed policy, \emph{compressed probing} switches 30 losing leaders to the winner. The winner's 32 original leaders continue comparisons against their shadow tags, while two matched leader pairs compare the compressed policies. In either probe mode, if the winning margin falls below half the entry threshold or the follower policy changes, the controller exits probe mode, restores all 64 leaders to their home policies, and suppresses counter updates for 4,096 leader accesses while they refill.

\emph{Changing the ceiling.}
A lower ceiling limits new page-table insertions to one five-way group under the 5-way policy or three uncompressed ways under the 3-way policy. Excess converted groups remain searchable for walk hits, avoiding an immediate cache-wide flush. When an application-data miss finds all unconverted ways occupied, the cache reclaims one excess group by evicting its resident lines (up to 64) and returning its five ways to data. If the write buffer fills before all lines are evicted, the group stays converted and subsequent data misses finish reclaiming it.

\subsection{Hardware Overhead}

\emph{Controller storage.}
Hashing the set index assigns leader roles without lookup storage. At each epoch's end, the controller updates per-core 7-bit miss weights from per-core access counters, off the critical path. The three 13-bit selection counters, per-core weights, epoch counters, and mode registers require 322 bits for eight cores or 147 bits for one core. Shadow tags require 6.4~KB; a lean variant using 12-bit partial tags and 16 leaders per compressed policy reduces this to 1.0~KB while matching performance within 0.3\%.

\emph{Implementation cost.}
To estimate hardware area, we model the 22~nm L2 arrays in CACTI~\cite{cacti} and synthesize our pipelined encoder, decoder, and slot-selection RTL with Synopsys Design Compiler in a 45~nm library. Each L2 uses a single encoder and decoder that accept one operation per cycle and, together with the slot-selection logic, take 24.4k NAND2-equivalent gates (18,712~$\mu$m$^2$ at 45~nm). Scaling that logic to 22~nm by a factor of 0.24 to 0.35 (accounting for $(22/45)^2$ feature scaling and interconnect overhead) adds 0.38 to 0.56\% to the area of a 1~MiB L2 cache, with each stage fitting inside a standard 64-bit ALU delay to support one-cycle fill encoding and two- or three-cycle decoding.

At 1~MiB and 4~MiB, the added tag and metadata bits increase L2 array area by 1.20\% and 1.74\% (leaving array access time unchanged at 1~MiB and adding 1.8\% at 4~MiB), the 6.4~KB shadow tag array adds 3.01\% and 0.85\% (or 0.80\% and 0.23\% for the lean 1.0~KB variant), and the controller, transition logic, multiplexers, and staging buffer add an estimated 0.2 to 0.5\% and 0.07 to 0.14\%. Altogether, LeafPack adds 4.8 to 5.3\% to the area of a 1~MiB L2 and 2.8 to 2.9\% to a 4~MiB L2 (or 2.6 to 3.1\% and 2.1 to 2.3\% with lean shadow tags) while increasing leakage power by 3.2\% and 2.4\%, excluding request sideband wires and place-and-route overhead.

\section{Implementation and Methodology}\label{sec:methodology}

\textit{Implementation and timing.}\label{sec:impl}
We implement LeafPack in roughly 2,000 lines of C++ in gem5~\cite{gem5,gem5v20.0+} (version 25.1) using its compressed-cache framework~\cite{gem5_compression_framework}, which supports multiple independently tagged logical lines per physical entry. We extend the shared L2 with the compressor, packed-set tag store, and set-dueling controller from Section~\ref{sec:leafpack} while leaving the TLBs, walk algorithm, and coherence protocol unmodified (walkers only tag page-table and leaf requests). We charge one cycle for fill encoding and add two or three cycles to L2 hits for packed or dense decoding, probing normal and packed sets in parallel assuming a banked tag array without port contention, and test a single-ported array in a separate model (Section~\ref{sec:eval-scaling}). We verify that disabling LeafPack reproduces unmodified gem5 statistics on every checkpoint, and we validate our tag store and coherence paths using runtime assertions, a 23.4M-operation reference-cache stress tester, and full-system page-remapping tests.

\textit{System and measurement.}
We run full-system Arm simulations under Linux~4.18 with 4~KiB base pages across two 4~GHz out-of-order core models, a small core (128-entry ROB with 12 L1 and 48 L2 MSHRs) and a server core (320-entry ROB with 48 L1 and 96 L2 MSHRs). Each core has private 32~KiB two-way L1 instruction and data caches, 64-entry L1 TLBs, a 1,280-entry five-way L2 TLB (caching upper-level partial translations so hits fetch only the leaf line), and up to 32 data walkers plus one instruction walker. The cores share a mostly exclusive 16-way L2 last-level cache with a 12-cycle access latency, a stride prefetcher, and 0.5~MiB of capacity per core (1, 2, and 4~MiB for two, four, and eight cores, while keeping 1~MiB for single programs to match the two-program pairs), backed by a dual-rank DDR3-1600 channel. We fast-forward on an atomic CPU to profiled phases and restore the same full-system checkpoint on the out-of-order CPUs for every configuration, warming up for 40 million cycles before simulating twenty 20-million-cycle intervals and reporting metrics over the last nineteen (380 million cycles, where including the first interval changes geometric-mean speedup by at most 0.2\%).

\textit{Workloads and validation.}
Table~\ref{tab:workloads} summarizes our unmodified benchmarks, where the \emph{translation} benchmarks allocate leaf page tables that exceed the 0.5~MiB per-core L2 budget by $9.0\times$ to $17.0\times$ (measuring total allocated leaf storage rather than the active subset reused in a single phase), while the \emph{co-runner} benchmarks have small leaf footprints and stress L2 data capacity. We run these programs alone and pinned one per core in two-, four-, and eight-program mixes,\footnote{Mix-1 combines bc, two pr, and 4~GiB SysBench. Mix-2 combines bc, XSBench, sssp, 4~GiB SysBench, two omnetpp, and two gcc. Mix-3 combines bc, bfs, pr, sssp, XSBench, 4~GiB SysBench, and two omnetpp.} separating them into an 18-workload development set (nine translation-heavy headline workloads and nine check workloads) used to design the controller and 20 randomly sampled held-out mixes (14 pairs and six four-program mixes) evaluated only after freezing the simulator binary.

\begin{table}[t]
\caption{\textbf{Benchmark page-table footprints.}}
\label{tab:workloads}
\centering
\footnotesize
\setlength{\tabcolsep}{2.5pt}
\resizebox{\columnwidth}{!}{%
\begin{tabular}{@{}lllrrrl@{}}
\toprule
Program & Suite & Input & Mapped & Leaves & Ratio & Role \\
 & & & (GiB) & (MiB) & ($\times$) & \\
\midrule
bc       & GAP~\cite{beamer2017gapbenchmarksuite} & U24 graph      & 2.46 & 4.94 & 9.9  & translation \\
pr       & GAP~\cite{beamer2017gapbenchmarksuite} & U24 graph      & 2.25 & 4.52 & 9.0  & translation \\
sssp     & GAP~\cite{beamer2017gapbenchmarksuite} & U24 graph      & 4.22 & 8.48 & 17.0 & translation \\
bfs      & GAP~\cite{beamer2017gapbenchmarksuite} & U24 graph      & 2.21 & 4.43 & 8.9  & moderate \\
XSBench  & XSBench~\cite{xsbench}                 & large grid     & 3.76 & 7.53 & 15.1 & translation \\
SysBench & SysBench~\cite{sysbench}               & 4\,GiB read    & 4.00 & 8.02 & 16.0 & translation \\
\midrule
SysBench & SysBench~\cite{sysbench}               & 512\,KiB read  & 0.003 & 0.02 & 0.05 & co-runner \\
omnetpp  & CPU2026~\cite{speccpu2026}             & reference      & 0.03 & 0.07 & 0.1  & co-runner \\
gcc      & CPU2026~\cite{speccpu2026}             & reference      & 0.33 & 0.73 & 1.5  & co-runner \\
\bottomrule
\end{tabular}%
}
\end{table}

\textit{Baselines and metrics.}
We evaluate the baseline cache with random replacement and, where noted, with 3-bit SRRIP or LRU. Both static uncompressed protection and LeafPack use the matching baseline replacement policy for application data while managing page-table lines with BRRIP~\cite{rrip} under a per-set way ceiling $W$. Under identical cache geometries, we compare the baseline against static uncompressed configurations ($W \in \{3, 5, 10, 15\}$), static compressed configurations ($W \in \{4, 5, 10, 15\}$, plus compressed $W = 3$ and uncompressed $W = 4$ on pr for the capacity study in Section~\ref{sec:eval-single}), LeafPack, and CoLT~\cite{colt} TLB coalescing. We also implement Victima~\cite{victima}, EWaH~\cite{ewah}, and xPTP~\cite{xptp} inside our shared L2 cache. In our Victima port, whenever its walk-cost predictor detects high translation pressure, the cache stores eight-translation clusters in dedicated L2 blocks under TLB-aware SRRIP and probes the L2 in parallel to skip page-table walks on a cluster hit. We do not charge Victima's 64~KiB predictor table or cluster headers against L2 capacity, and we report its best per-set way cap from $N\in\{4,8,16\}$ on each workload. To model EWaH and xPTP, we add a priority bit to each L2 cache line with no access-latency penalty. Under EWaH, the cache protects page-table walker fills from eviction 99\% of the time while the requesting core experiences frequent TLB misses, and it never allows one core's data accesses to evict another core's protected page-table lines. Under xPTP, which relies on LRU recency order, the cache evicts the LRU data line whenever one sits within eight positions of the LRU tail (falling back to the true LRU line otherwise) and turns off page-table priority when translation pressure is low.

We measure single-program throughput using IPC and multiprogrammed throughput using harmonic speedup~\cite{luo2001hs} relative to the same mix on the baseline, $\mathrm{HS}=n/\sum_{i}(\mathrm{IPC}_i^{\mathrm{base}}/\mathrm{IPC}_i^{\mathrm{config}})$.
Harmonic speedup reduces to the IPC ratio for one program and penalizes any policy that accelerates one program by starving a co-runner of data capacity. To explain the resulting trends, we also track page-walk misses, data misses, and page-table cache occupancy.

\section{Evaluation}\label{sec:evaluation}

Unless noted otherwise, paired numbers report the small and server cores using the cache sizes from Section~\ref{sec:methodology}.

\textit{Interpreting the comparisons.}
We evaluate LeafPack against three reference points to separate its total speedup into the benefit of protecting page-table lines, the benefit of compressing them, and the cost of online policy selection.
\begin{itemize}[topsep=0pt, itemsep=1pt, leftmargin=*]
  \item \textbf{Baseline.} We use the shared L2 cache defined in the introduction to measure LeafPack's overall harmonic speedup (\emph{HS}).
  \item \textbf{Best static uncompressed.} We pick the highest-performing configuration offline for each workload among the baseline and static uncompressed protection ($W \in \{3, 5, 10, 15\}$) to isolate the gain of leaf compression beyond page-table protection alone.
  \item \textbf{Best static compressed.} We pick the highest-performing static compressed configuration offline for each workload ($W \in \{4, 5, 10, 15\}$) to measure the cost of restricting our runtime controller to three candidate policies and selecting among them online.
\end{itemize}

\subsection{Capacity Savings for Single Programs}\label{sec:eval-single}

To assess the headroom available from leaf compression before evaluating our runtime controller, we first test static way ceilings. While our adaptive controller is restricted to the three candidate policies from Section~\ref{sec:controller} (five or fifteen compressed ways, or three uncompressed ways), we also allow the static sweep to form \emph{partial} groups in which one way holds metadata and the remaining ways hold compressed payloads (forming three 80-byte slots across four ways at $W{=}4$ with 16 metadata bytes unused, or two slots across three ways at $W{=}3$ with 32 bytes unused). Our reported compressed occupancy charges every converted way, including metadata, unused bytes, and empty slots, while uncompressed occupancy counts resident 64-byte lines.

\begin{figure}[b]
\centering
\pgfplotsset{sgbar/.style={lpbase, scale only axis, clip=false, width=3.3cm, height=0.62cm, xmin=0.45, xmax=5.55,
  ybar, bar width=0.34, bar shift=0pt, xmajorgrids=false, ymajorgrids=true, axis x line*=bottom, axis y line*=left,
  every axis plot/.append style={draw=white, line width=0.25pt}, xtick={1,...,5}, xticklabels={}, xtick style={draw=none},
  yticklabel style={font=\lpLeg}, x tick label style={font=\lpLeg, inner sep=1pt}},
  sgLab/.style={xticklabels={3,4,5,10,15}},
  sgtri/.style={only marks, mark=triangle*, mark size=2.2pt, black, mark options={rotate=180, fill=black, draw=white, line width=0.3pt}, forget plot}}
\tikzset{sgRaw/.style={fill=lpVerm!30, draw=lpVerm!80!black, line width=0.3pt, lpHatch}, sgLeaf/.style={fill=lpBlue, draw=lpBlue!70!black, line width=0.3pt}}
\newcommand{\sgpos}{(\thisrow{W}==3 ? 1 : (\thisrow{W}==4 ? 2 : (\thisrow{W}==5 ? 3 : (\thisrow{W}==10 ? 4 : 5))))}
\newcommand{\sgpanel}[3]{%
\begin{axis}[sgbar, #3]
\addplot[sgRaw]  table[x expr={\sgpos-0.18}, y expr={\thisrow{W}==0 ? nan : (\thisrow{#2}-1)*100}] {single-#1-raw.dat};
\addplot[sgLeaf] table[x expr={\sgpos+0.18}, y expr={\thisrow{W}==0 ? nan : (\thisrow{#2}-1)*100}] {single-#1-leaf.dat};
\draw[black!60, line width=0.4pt] (axis cs:0.45,0) -- (axis cs:5.55,0);
\end{axis}}
\begin{tikzpicture}
\sgpanel{bc}{small}{ymin=-5, ymax=38, ytick={0,15,30}}
\node[font=\lpLab\bfseries, anchor=south] at (1.65cm,0.75cm) {Small core};
\node[font=\lpLab\bfseries, anchor=south] at (5.4cm,0.75cm) {Server core};
\begin{axis}[sgbar, ymin=-5, ymax=38, hide axis]\addplot[sgtri] coordinates {(2.18,31.47+5)};\end{axis}
\sgpanel{bc}{server}{at={(3.75cm,0)}, ymin=-5, ymax=38, ytick={0,15,30}, yticklabels={}}
\begin{axis}[sgbar, at={(3.75cm,0)}, ymin=-5, ymax=38, hide axis]\addplot[sgtri] coordinates {(2.18,28.34+5)};\end{axis}
\sgpanel{pr}{small}{at={(0,-0.94cm)}, ymin=0, ymax=22, ytick={0,10,20}}
\begin{axis}[sgbar, at={(0,-0.94cm)}, ymin=0, ymax=22, hide axis]\addplot[sgtri] coordinates {(2.18,12.81+2.9)};\end{axis}
\sgpanel{pr}{server}{at={(3.75cm,-0.94cm)}, ymin=0, ymax=22, ytick={0,10,20}, yticklabels={}}
\begin{axis}[sgbar, at={(3.75cm,-0.94cm)}, ymin=0, ymax=22, hide axis]\addplot[sgtri] coordinates {(2.18,18.20+2.9)};\end{axis}
\sgpanel{sssp}{small}{at={(0,-1.88cm)}, ymin=0, ymax=22, ytick={0,10,20}}
\begin{axis}[sgbar, at={(0,-1.88cm)}, ymin=0, ymax=22, hide axis]\addplot[sgtri] coordinates {(3.18,19.70+2.8)};\end{axis}
\sgpanel{sssp}{server}{at={(3.75cm,-1.88cm)}, ymin=0, ymax=22, ytick={0,10,20}, yticklabels={}}
\begin{axis}[sgbar, at={(3.75cm,-1.88cm)}, ymin=0, ymax=22, hide axis]\addplot[sgtri] coordinates {(3.18,18.71+2.8)};\end{axis}
\sgpanel{XSBench}{small}{at={(0,-2.82cm)}, ymin=0, ymax=38, ytick={0,15,30}}
\begin{axis}[sgbar, at={(0,-2.82cm)}, ymin=0, ymax=38, hide axis]\addplot[sgtri] coordinates {(2.18,30.93+5)};\end{axis}
\sgpanel{XSBench}{server}{at={(3.75cm,-2.82cm)}, ymin=0, ymax=38, ytick={0,15,30}, yticklabels={}}
\begin{axis}[sgbar, at={(3.75cm,-2.82cm)}, ymin=0, ymax=38, hide axis]\addplot[sgtri] coordinates {(2.18,32.81+5)};\end{axis}
\sgpanel{SysBench}{small}{at={(0,-3.76cm)}, ymin=0, ymax=75, ytick={0,35,70}, sgLab}
\begin{axis}[sgbar, at={(0,-3.76cm)}, ymin=0, ymax=75, hide axis]\addplot[sgtri] coordinates {(2.18,6.52+8)};\end{axis}
\sgpanel{SysBench}{server}{at={(3.75cm,-3.76cm)}, ymin=0, ymax=75, ytick={0,35,70}, yticklabels={}, sgLab}
\begin{axis}[sgbar, at={(3.75cm,-3.76cm)}, ymin=0, ymax=75, hide axis]\addplot[sgtri] coordinates {(2.18,6.42+8)};\end{axis}
\foreach \n/\y in {bc/0, pr/-0.94, sssp/-1.88, XSBench/-2.82, SysBench 4\,GiB/-3.76}
  \node[font=\lpLeg\bfseries, anchor=south west, inner sep=0.5pt] at (0.0cm,\y cm+0.66cm) {\n};
\node[font=\lpLab, anchor=north] at (3.5cm,-4.08cm) {Way budget $W$};
\node[font=\lpLab, rotate=90, anchor=south] at (-0.55cm,-1.55cm) {IPC gain over baseline (\%)};
\end{tikzpicture}
\\[0.3mm]{\lpLeg\tikz\draw[sgRaw] (0,0) rectangle (0.3,0.18);~Static uncompressed\hspace{3mm}\tikz\draw[sgLeaf] (0,0) rectangle (0.3,0.18);~Static compressed\hspace{3mm}\tikz\fill (0,0.15) -- (0.18,0.15) -- (0.09,0) -- cycle;~Matching $W$}
\caption{\textbf{Single-program capacity savings.} Triangles mark the smallest compressed $W$ within 0.1\% of the best uncompressed IPC.}
\label{fig:single}
\end{figure}
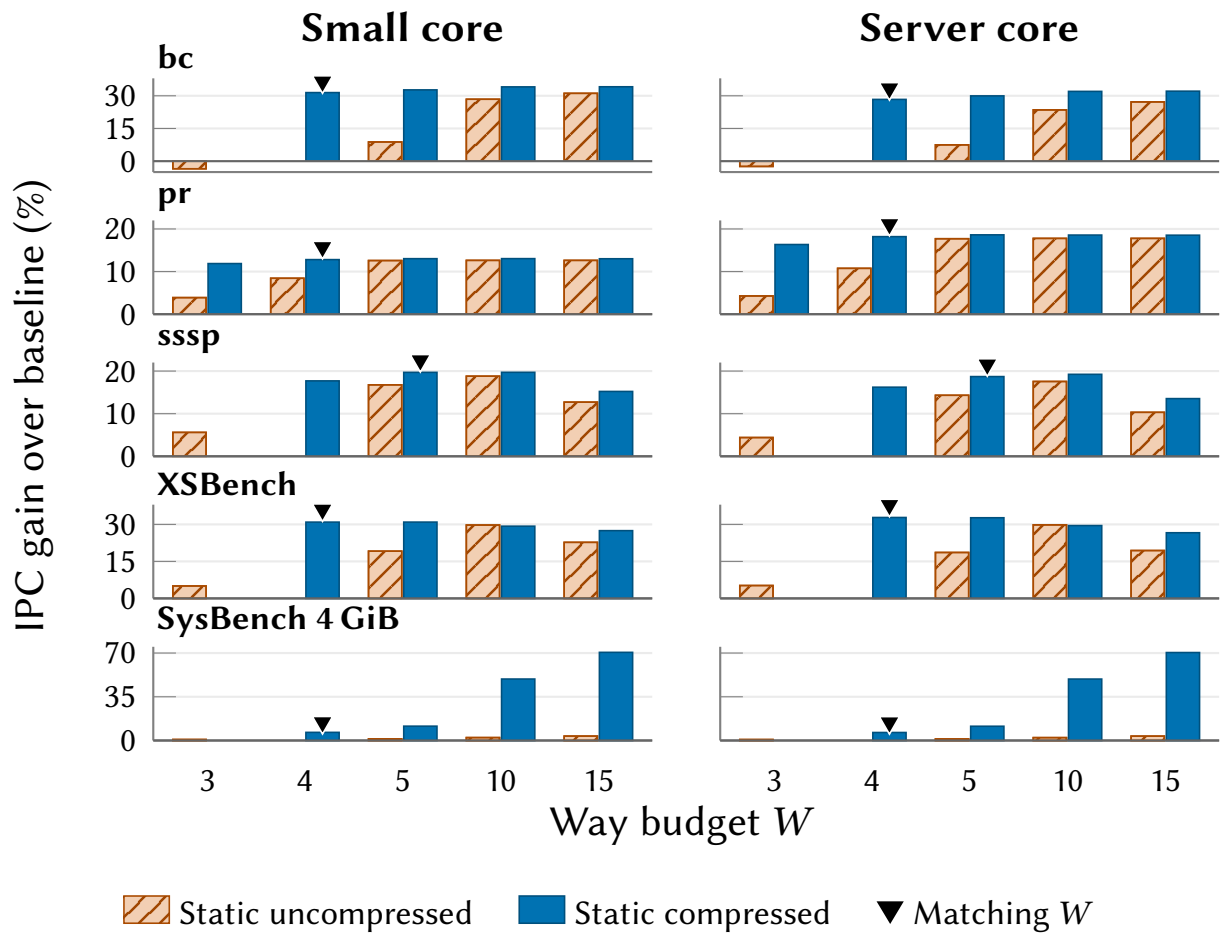

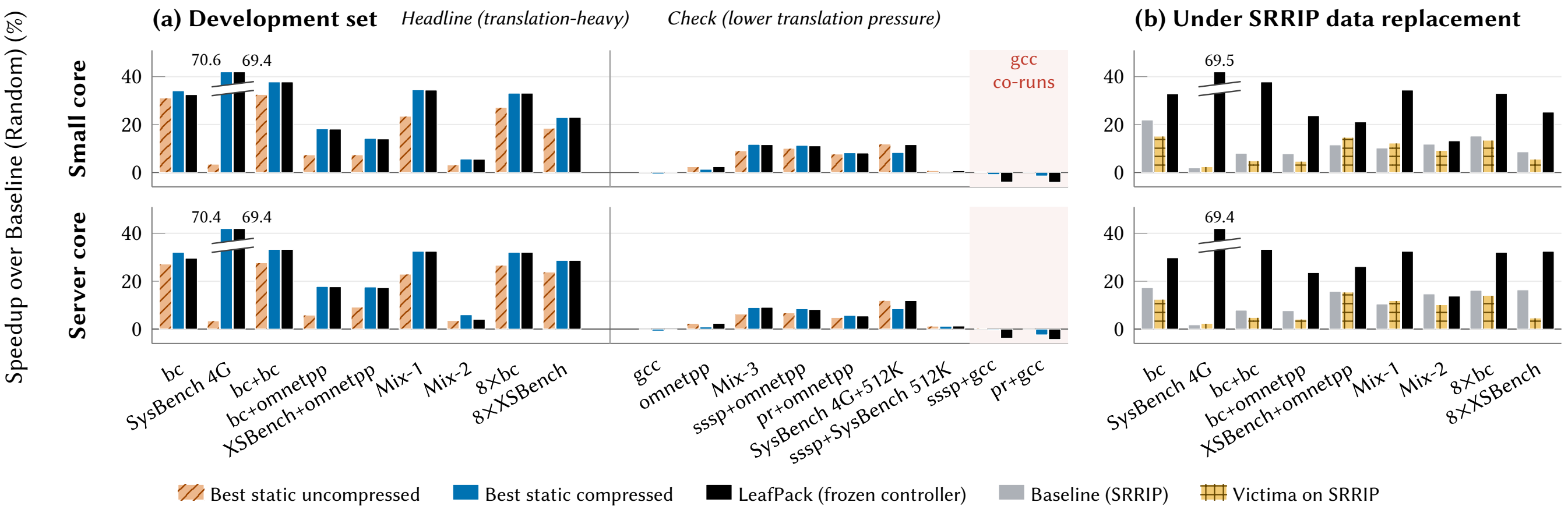
\begin{figure*}[t]
\centering
\pgfplotsset{dvbar/.style={lpbase, scale only axis, clip=false, height=1.55cm, ymin=-6, ymax=51, ytick={0,20,40},
  ybar, bar width=0.25, bar shift=0pt, xmajorgrids=false, ymajorgrids=true, axis x line*=bottom, axis y line*=left,
  every axis plot/.append style={draw=white, line width=0.25pt}, xtick=\empty, xtick style={draw=none}},
  dvA/.style={dvbar, width=10.4cm, xmin=0.45, xmax=19.55}, dvB/.style={dvbar, width=4.85cm, xmin=0.45, xmax=9.55},
  dvLabelsA/.style={xtick={1,...,9,11,12,...,19}, x tick label style={rotate=30, anchor=north east, inner sep=0.5pt},
    xticklabels={bc, SysBench 4G, bc+bc, bc+omnetpp, XSBench+omnetpp, Mix-1, Mix-2, 8$\times$bc, 8$\times$XSBench,
    gcc, omnetpp, Mix-3, sssp+omnetpp, pr+omnetpp, SysBench 4G+512K, sssp+SysBench 512K, sssp+gcc, pr+gcc}},
  dvLabelsB/.style={xtick={1,...,9}, x tick label style={rotate=30, anchor=north east, inner sep=0.5pt},
    xticklabels={bc, SysBench 4G, bc+bc, bc+omnetpp, XSBench+omnetpp, Mix-1, Mix-2, 8$\times$bc, 8$\times$XSBench}},
  dvClip/.style={only marks, mark=none, draw=none, forget plot, point meta=explicit,
    nodes near coords={\pgfmathprintnumber[fixed, precision=1, assume math mode=true]{\pgfplotspointmeta}},
    nodes near coords style={font=\fontsize{6}{6}\selectfont\sffamily, anchor=center, inner sep=0pt}}}
\tikzset{dvRaw/.style={fill=lpVerm!45, lpHatch}, dvLeaf/.style={fill=lpBlue}, dvCtl/.style={fill=black},
  dvSr/.style={fill=lpGrey!70}, dvVic/.style={fill=lpVic!55, postaction={pattern={Hatch[distance=2.8pt, line width=0.4pt]}, pattern color=lpVic!45!black}},
  dvBreak/.pic={\draw[white, line width=2.6pt] (-0.24cm,-0.03cm) -- (0.24cm,0.03cm); \draw[black!75, line width=0.45pt] (-0.24cm,-0.08cm) -- (0.24cm,-0.02cm) (-0.24cm,0.02cm) -- (0.24cm,0.08cm);}}
\newcommand{\dvrowA}[3]{%
\begin{axis}[dvA, #2]
\fill[lpRed!6] (axis cs:17.5,-6) rectangle (axis cs:19.55,51);
\draw[black!35, line width=0.5pt] (axis cs:10,-6) -- (axis cs:10,51);
\draw[black!60, line width=0.4pt] (axis cs:0.45,0) -- (axis cs:19.55,0);
\addplot[dvRaw]  table[x expr=\thisrow{x}-0.27, y expr={min((\thisrow{raw#1}-1)*100,42)}] {dev.dat};
\addplot[dvLeaf] table[x expr=\thisrow{x},      y expr={min((\thisrow{ora#1}-1)*100,42)}] {dev.dat};
\addplot[dvCtl]  table[x expr=\thisrow{x}+0.27, y expr={min((\thisrow{ctl#1}-1)*100,42)}] {dev.dat};
\addplot[dvClip] table[x expr=\thisrow{x}-0.42,      y expr={\thisrow{ora#1}>1.42 ? 47 : nan}, meta expr={(\thisrow{ora#1}-1)*100}] {dev.dat};
\addplot[dvClip] table[x expr=\thisrow{x}+0.64, y expr={\thisrow{ctl#1}>1.42 ? 47 : nan}, meta expr={(\thisrow{ctl#1}-1)*100}] {dev.dat};
\path (axis cs:2.135,35) pic {dvBreak};
\node[font=\lpLab\bfseries, anchor=south, rotate=90] at (rel axis cs:-0.062,0.5) {#3};
\end{axis}}
\newcommand{\dvrowB}[2]{%
\begin{axis}[dvB, #2]
\draw[black!60, line width=0.4pt] (axis cs:0.45,0) -- (axis cs:9.55,0);
\addplot[dvSr]  table[x expr=\thisrow{x}-0.27, y expr={min((\thisrow{plain#1}-1)*100,42)}] {srrip.dat};
\addplot[dvVic] table[x expr=\thisrow{x},      y expr={min((\thisrow{vic#1}-1)*100,42)}] {srrip.dat};
\addplot[dvCtl] table[x expr=\thisrow{x}+0.27, y expr={min((\thisrow{lp#1}-1)*100,42)}] {srrip.dat};
\addplot[dvClip] table[x expr=\thisrow{x}+0.27, y expr={\thisrow{lp#1}>1.42 ? 47 : nan}, meta expr={(\thisrow{lp#1}-1)*100}] {srrip.dat};
\path (axis cs:2.27,35) pic {dvBreak};
\end{axis}}
\begin{tikzpicture}
\dvrowA{S}{}{Small core}
\node[font=\lpLab\bfseries, anchor=south west, inner sep=0pt] at (0,1.75cm) {(a) Development set};
\node[font=\lpLeg\itshape, anchor=south east, inner sep=1pt] at (5.45cm,1.75cm) {Headline (translation-heavy)};
\node[font=\lpLeg\itshape, anchor=south west, inner sep=1pt] at (5.8cm,1.75cm) {Check (lower translation pressure)};
\node[font=\lpLeg, text=lpRed, anchor=north, inner sep=1pt, align=center] at (9.9cm,1.53cm) {gcc\\co-runs};
\dvrowA{V}{at={(0,-1.78cm)}, dvLabelsA}{Server core}
\node[font=\lpLab, rotate=90, anchor=south] at (-1.3cm,-0.12cm) {Speedup over Baseline (Random) (\%)};
\begin{scope}[xshift=11.15cm]
\dvrowB{S}{}
\node[font=\lpLab\bfseries, anchor=south west, inner sep=0pt] at (0,1.75cm) {(b) Under SRRIP data replacement};
\dvrowB{V}{at={(0,-1.78cm)}, dvLabelsB}
\end{scope}
\end{tikzpicture}
\\[-0.5mm]{\lpLeg
\tikz\draw[dvRaw, draw=white] (0,0) rectangle (0.3,0.18);~Best static uncompressed\hspace{3mm}
\tikz\draw[dvLeaf, draw=white] (0,0) rectangle (0.3,0.18);~Best static compressed\hspace{3mm}
\tikz\draw[dvCtl, draw=white] (0,0) rectangle (0.3,0.18);~LeafPack (frozen controller)\hspace{3mm}
\tikz\draw[dvSr, draw=white] (0,0) rectangle (0.3,0.18);~Baseline (SRRIP)\hspace{3mm}
\tikz\draw[dvVic, draw=white] (0,0) rectangle (0.3,0.18);~Victima on SRRIP}
\caption{Harmonic speedup gain over our random-replacement baseline across the development set (a) and under SRRIP data replacement (b) on the small (top) and server (bottom) cores. LeafPack trails the SRRIP baseline only on server-core Mix-2.}
\label{fig:dev}
\end{figure*}

As shown in Fig.~\ref{fig:single}, compressing leaf lines matches the IPC of the best static uncompressed configuration while consuming 50 to 73\% less physical L2 capacity (saving 73\% on bc and 4~GiB SysBench, 64\% on pr, 60\% on XSBench, and 50\% on sssp). Under an equal five-way ceiling, LeafPack improves IPC whenever the active leaf working set overflows the uncompressed ways, gaining 22.0\% on bc and 9.9\% on XSBench. On bc, four compressed ways already match fifteen uncompressed ways on both the small and server cores, so additional page-table ways provide little standalone benefit and instead remain available to co-runners. By contrast, on 4~GiB SysBench, the active 8~MiB leaf working set is too large to fit in the 1~MiB L2 even at fifteen uncompressed ways, so leaf compression continues to expand translation coverage all the way to $W{=}15$ and raises IPC by $1.65\times$ over fifteen uncompressed ways ($1.71\times$ over the baseline).

\subsection{Adaptive Controller}\label{sec:eval-dev}

Figure~\ref{fig:dev}a evaluates our frozen runtime controller across all 18 development workloads. On the nine translation-heavy headline workloads (left), LeafPack improves geometric-mean harmonic speedup over the baseline by 28.7\% on the small core and 28.4\% on the server core, outperforming the best static uncompressed configuration on both cores and staying within 1\% of best static compressed on every run except single-program bc and server-core Mix-2. On the nine check workloads (right), our controller never falls more than 0.5\% below the best static uncompressed configuration except on the two co-runs with gcc (Section~\ref{sec:discussion}).

\textit{Adaptation and oracle gaps.}
Figure~\ref{fig:cases} illustrates how our runtime controller converges on each of the three candidate policies across three representative workloads. When two instances of bc share the cache (Fig.~\ref{fig:cases}a), the 15-way compressed policy A beats the 5-way policy B by 7.5 to 9.7\% in every interval and the controller tracks policy A from the start. When bc instead co-runs with data-hungry omnetpp (Fig.~\ref{fig:cases}b), policy B beats policy A by 15.4\% by preserving data capacity for omnetpp, and the controller follows policy B across the window. When omnetpp runs alone (Fig.~\ref{fig:cases}c), its small leaf footprint fits inside 3 uncompressed ways so policy C beats both compressed policies by 2\%, and the controller selects policy C and enters probe mode within three intervals so only four physical leader sets continue running the compressed policies.

The controller also adapts when shared-cache demand shifts mid-run, switching follower sets between policies A and B across pr's traversal phases on pr+omnetpp. To test a sharper phase change where no single static policy is optimal for the entire run, we pause the data-sensitive co-runner in 4~GiB+512~KiB SysBench for ten intervals before resuming it. The controller shifts from policy C to the 15-way policy A while 4~GiB SysBench runs alone and then drops to the 5-way policy B once the co-runner resumes, outperforming every static policy over the full run (1.124 versus 1.111 harmonic speedup, while trailing the best static choice by 2.3\% and 3.4\% in the same test on pr+gcc).

Among our headline workloads, standalone bc and server-core Mix-2 show the two largest gaps to the best static compressed configuration for different reasons. On standalone bc, nearly tied leader-set miss counts cause the controller to pick five compressed ways, leaving a 1.2 to 1.9\% gap below fifteen ways, whereas on server-core Mix-2 the best static setting uses a four-way partial group outside our candidate set, where the best candidate reaches 0.985 and our controller reaches 0.982.

\colorlet{pk}{lpBlue}
\definecolor{dn}{HTML}{6A4C93}
\colorlet{lvD}{lpVerm}
\begin{figure}[b]
\centering
\pgfplotsset{caseax/.style={scale only axis, width=0.355\columnwidth, height=1.3cm, font=\scriptsize\sffamily,
  title style={font=\scriptsize\sffamily\bfseries, yshift=-1.5mm}, xmin=0.5, xmax=19.5, xtick={1,10,19},
  grid=major, grid style={black!8}, axis line style={black!50}, tick style={black!50},
  yticklabel style={/pgf/number format/fixed, /pgf/number format/precision=2},
  legend style={font=\scriptsize\sffamily, draw=none, legend columns=1, row sep=-1pt, legend cell align=left}}}
\begin{tikzpicture}[baseline=(current axis.south)]
\begin{axis}[caseax, title={(a) bc+bc, small core}, ymin=0.92, ymax=1.44, ytick={1,1.2,1.4}, ylabel={HS over baseline}, ylabel style={yshift=-1mm}]
\addplot[pk, line width=0.9pt] coordinates {(1,1.3631) (2,1.3674) (3,1.3830) (4,1.3662) (5,1.3729) (6,1.3864) (7,1.3772) (8,1.3765) (9,1.3869) (10,1.3693) (11,1.3783) (12,1.3832) (13,1.3850) (14,1.3805) (15,1.3782) (16,1.3826) (17,1.3896) (18,1.3813) (19,1.3832)};
\addplot[dn, line width=0.9pt, lpdash] coordinates {(1,1.2495) (2,1.2673) (3,1.2680) (4,1.2578) (5,1.2633) (6,1.2663) (7,1.2642) (8,1.2628) (9,1.2714) (10,1.2740) (11,1.2602) (12,1.2659) (13,1.2663) (14,1.2619) (15,1.2692) (16,1.2598) (17,1.2697) (18,1.2669) (19,1.2608)};
\addplot[lvD, line width=0.9pt, dash pattern=on 4pt off 1.5pt on 1pt off 1.5pt] coordinates {(1,0.9726) (2,0.9537) (3,0.9651) (4,0.9659) (5,0.9664) (6,0.9666) (7,0.9638) (8,0.9641) (9,0.9658) (10,0.9579) (11,0.9757) (12,0.9561) (13,0.9655) (14,0.9725) (15,0.9566) (16,0.9649) (17,0.9564) (18,0.9748) (19,0.9668)};
\addplot[black, only marks, mark=*, mark size=2.2pt, mark options={fill=white, line width=0.8pt}] coordinates {(1,1.3618) (2,1.3677) (3,1.3762) (4,1.3734) (5,1.3719) (6,1.3872) (7,1.3768) (8,1.3761) (9,1.3883) (10,1.3689) (11,1.3786) (12,1.3810) (13,1.3847) (14,1.3803) (15,1.3786) (16,1.3830) (17,1.3898) (18,1.3793) (19,1.3780)};
\node[font=\scriptsize\sffamily, anchor=south east, fill=white, fill opacity=0.85, text opacity=1, inner sep=1pt] at (rel axis cs:0.98,0.32) {Controller selects A};
\end{axis}
\end{tikzpicture}\hfill
\begin{tikzpicture}[baseline=(current axis.south)]
\begin{axis}[caseax, title={(b) bc+omnetpp, small core}, ymin=0.98, ymax=1.22, ytick={1,1.1,1.2}]
\addplot[pk, line width=0.9pt] coordinates {(1,1.0235) (2,1.0369) (3,1.0427) (4,1.0267) (5,1.0226) (6,1.0225) (7,1.0211) (8,1.0182) (9,1.0160) (10,1.0271) (11,1.0283) (12,1.0140) (13,1.0284) (14,1.0214) (15,1.0269) (16,1.0213) (17,1.0177) (18,1.0073) (19,1.0214)};
\addplot[dn, line width=0.9pt, lpdash] coordinates {(1,1.1773) (2,1.1750) (3,1.1813) (4,1.1780) (5,1.1787) (6,1.1849) (7,1.1774) (8,1.1838) (9,1.1793) (10,1.1831) (11,1.1860) (12,1.1758) (13,1.1817) (14,1.1855) (15,1.1842) (16,1.1852) (17,1.1853) (18,1.1860) (19,1.1804)};
\addplot[lvD, line width=0.9pt, dash pattern=on 4pt off 1.5pt on 1pt off 1.5pt] coordinates {(1,1.0081) (2,1.0093) (3,1.0145) (4,1.0124) (5,1.0113) (6,1.0136) (7,1.0111) (8,1.0085) (9,1.0127) (10,1.0111) (11,1.0158) (12,1.0050) (13,1.0109) (14,1.0156) (15,1.0096) (16,1.0119) (17,1.0158) (18,1.0093) (19,1.0094)};
\addplot[black, only marks, mark=*, mark size=2.2pt, mark options={fill=white, line width=0.8pt}] coordinates {(1,1.1749) (2,1.1752) (3,1.1834) (4,1.1763) (5,1.1792) (6,1.1861) (7,1.1785) (8,1.1852) (9,1.1810) (10,1.1801) (11,1.1865) (12,1.1753) (13,1.1819) (14,1.1842) (15,1.1831) (16,1.1852) (17,1.1835) (18,1.1847) (19,1.1801)};
\node[font=\scriptsize\sffamily, anchor=south east, fill=white, fill opacity=0.85, text opacity=1, inner sep=1pt] at (rel axis cs:0.98,0.32) {Controller selects B};
\end{axis}
\end{tikzpicture}\\[0.5mm]
\begin{tikzpicture}[baseline=(current axis.south)]
\begin{axis}[caseax, legend to name=caselegend, title={(c) omnetpp, server core}, ymin=0.99, ymax=1.045, ytick={1,1.02,1.04}, ylabel={HS over baseline}, ylabel style={yshift=-1mm}, xlabel={Interval (20M cycles)}, xlabel style={yshift=1.5mm}]
\addplot[pk, line width=0.9pt] coordinates {(1,1.0065) (2,1.0033) (3,1.0074) (4,1.0028) (5,1.0065) (6,1.0011) (7,1.0044) (8,1.0037) (9,0.9991) (10,1.0014) (11,1.0072) (12,1.0046) (13,1.0067) (14,0.9982) (15,1.0021) (16,1.0079) (17,1.0043) (18,1.0002) (19,1.0063)};
\addplot[dn, line width=0.9pt, lpdash] coordinates {(1,1.0064) (2,1.0008) (3,1.0051) (4,1.0039) (5,1.0057) (6,1.0055) (7,1.0038) (8,0.9999) (9,0.9987) (10,1.0047) (11,1.0053) (12,1.0062) (13,1.0091) (14,0.9989) (15,1.0039) (16,1.0050) (17,1.0053) (18,1.0001) (19,1.0047)};
\addplot[lvD, line width=0.9pt, dash pattern=on 4pt off 1.5pt on 1pt off 1.5pt] coordinates {(1,1.0248) (2,1.0211) (3,1.0085) (4,1.0316) (5,1.0278) (6,1.0227) (7,1.0264) (8,1.0185) (9,1.0161) (10,1.0194) (11,1.0179) (12,1.0249) (13,1.0233) (14,1.0175) (15,1.0085) (16,1.0314) (17,1.0244) (18,1.0166) (19,1.0339)};
\addplot[black, only marks, mark=*, mark size=2.2pt, mark options={fill=white, line width=0.8pt}] coordinates {(1,1.0229) (2,1.0214) (3,1.0264) (4,1.0259) (5,1.0260) (6,1.0220) (7,1.0273) (8,1.0219) (9,1.0183) (10,1.0234) (11,1.0199) (12,1.0254) (13,1.0294) (14,1.0210) (15,1.0149) (16,1.0350) (17,1.0267) (18,1.0161) (19,1.0376)};
\legend{A: 15 compressed ways, B: 5 compressed ways, C: 3 uncompressed ways, Controller}
\node[font=\scriptsize\sffamily, anchor=north west, fill=white, fill opacity=0.85, text opacity=1, inner sep=1pt] at (rel axis cs:0.02,0.98) {Controller selects C};
\end{axis}
\end{tikzpicture}\hfill
\begin{minipage}[b]{0.40\columnwidth}\centering\pgfplotslegendfromname{caselegend}\par\vspace{4.3mm}\end{minipage}
\caption{Runtime policy adaptation.}
\label{fig:cases}
\end{figure}
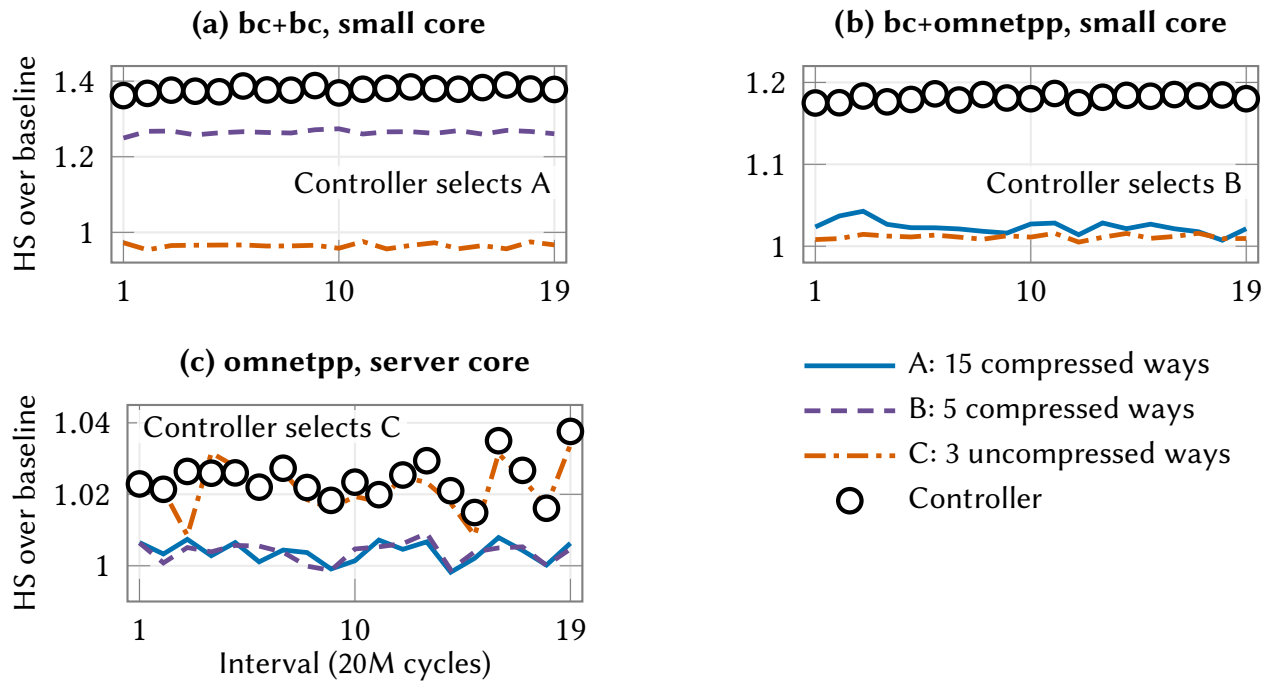

\textit{Controller ablations.}
We next ablate our controller's design choices to see how each mechanism contributes. First, evaluating the 3-way uncompressed policy in shadow tags rather than dedicated physical leader sets avoids forcing translation-heavy accesses through sets whose three ways cannot hold a large leaf working set, gaining 2.0\% on bc+bc. Second, entering compressed probe mode cuts the ongoing miss cost of keeping all 32 leader sets on a losing compressed ceiling, lifting 4~GiB SysBench by 2.0\% and the omnetpp pairs by 0.3 to 0.9\%. Finally, because compressed slots fill incrementally over time, our warm-up gate prevents early cold misses in compressed leader sets from locking follower sets onto the uncompressed policy during warm-up, raising harmonic speedup on server-core $8\times$XSBench from 0.992 to 1.287.

\textit{Selection accuracy and fairness.}
Across the full cache, our per-core weighted miss metric picks the candidate with the highest harmonic speedup in 35 of 36 workload-core cases (with the sole exception on server-core pr+omnetpp being a 0.03\% tie), though leader-set sampling can cost 3 to 4\% in harmonic speedup when candidates differ by under 2\% in weighted misses. To separate online selection loss from the cost of restricting the controller to three candidates, we compare against all eight static configurations from Section~\ref{sec:methodology}. Across the 18 development workloads on both cores, the best of our three candidates reaches 1.000 and 0.999 of the best among all eight static configurations (candidate-set loss), and our online controller reaches 0.994 of its best candidate (selection loss), with most of that gap coming from the two gcc pairs (0.964 and 0.960, where candidate C is the best candidate). Outside the development set, fragmented XSBench experiences both selection and candidate-set loss (reaching 0.911 and 0.919 of its best candidate, while its best candidate reaches 0.946 and 0.928 of the 10-way static ceiling in Section~\ref{sec:eval-scaling}). Finally, even when a multiprogrammed mix improves in overall harmonic speedup, individual co-runners within the mix can still experience a slowdown~\cite{kim2004fairsharing} when page-table ways displace their data (slowing by 0.9 to 16\% in mixes that improve overall, and by up to 24\% for gcc when paired with sssp, as analyzed in Section~\ref{sec:discussion}).


\subsection{Where the Gain Comes From}\label{sec:eval-breakdown}

Across all 18 development workloads, LeafPack improves harmonic speedup by 15.7\% and 15.1\% over the baseline. While protecting page-table lines accounts for most of that gain, LeafPack still leads the best static uncompressed configuration by 4.7\% and 4.8\%, which we break down in the rest of this section.

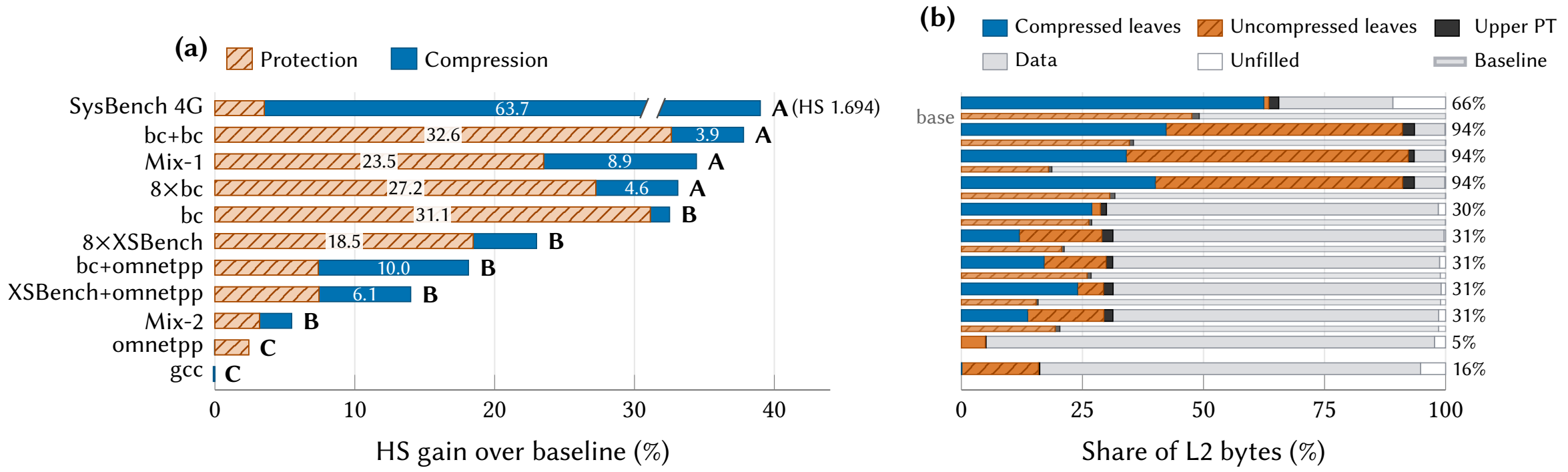
\begin{figure*}[t]
\centering
\pgfplotsset{bdax/.style={lpbase, scale only axis, clip=false, height=2.95cm, ymin=-0.6, ymax=10.6, ytick={0,...,10},
  ymajorgrids=false, xmajorgrids=true, grid style={black!9}, axis x line*=bottom, y axis line style={draw=none}, major y tick style={draw=none}, yticklabels={},
  legend style={font=\lpLeg, at={(0,1.02)}, anchor=south west, /tikz/every even column/.append style={column sep=2.5mm}},
  legend image code/.code={\draw[#1] (0cm,-0.09cm) rectangle (0.3cm,0.09cm);}},
  bdA/.style={bdax, width=6.1cm, xmin=0, xmax=44, yticklabels from table={breakdown.dat}{name}}, bdB/.style={bdax, width=4.8cm, xmin=0, xmax=100, at={(7.4cm,0)}},
  bdLab/.style={only marks, mark=none, forget plot, point meta=explicit,
    nodes near coords={\pgfmathprintnumber[fixed, precision=1, fixed zerofill, assume math mode=true]{\pgfplotspointmeta}}, nodes near coords style={anchor=center, inner sep=0pt}}}
\tikzset{bdRaw/.style={fill=lpVerm!30, draw=lpVerm!80!black, line width=0.3pt, lpHatch}, bdLeaf/.style={fill=lpBlue, draw=lpBlue!70!black, line width=0.3pt},
  bdRawS/.style={fill=lpVerm!80, draw=lpVerm!80!black, line width=0.3pt}, bdUp/.style={fill=black!80, draw=black, line width=0.3pt}, bdData/.style={fill=lpGrey!30, draw=lpGrey, line width=0.3pt},
  bdEmpty/.style={fill=white, draw=lpGrey, line width=0.3pt}}
\begin{tikzpicture}
\begin{axis}[bdA, xtick={0,10,20,30,40}, xlabel={HS gain over baseline (\%)}, xbar stacked, bar width=0.56, legend columns=2]
\addplot[bdRaw] table[y=y, x expr={(\thisrow{raw}-1)*100}] {breakdown.dat};
\addplot[bdLeaf] table[y=y, x expr={min((\thisrow{ctl}-\thisrow{raw})*100, 39-(\thisrow{raw}-1)*100)}] {breakdown.dat};
\legend{Protection, Compression}
\end{axis}
\begin{axis}[bdA, hide axis, nodes near coords, point meta=explicit symbolic,
  nodes near coords style={font=\lpTick\bfseries, anchor=west, inner sep=1.5pt}]
\addplot[only marks, mark=none] table[y=y, x expr={min((\thisrow{ctl}-1)*100, 39)+0.3}, meta=pol] {breakdown.dat};
\addplot[bdLab, nodes near coords style={font=\fontsize{6}{6}\selectfont\sffamily, fill=white, fill opacity=0.9, text opacity=1, inner sep=0.4pt}]
  table[y=y, x expr={(\thisrow{raw}-1)*50}, y expr={(\thisrow{raw}-1)*100>=9 ? \thisrow{y} : nan}, meta expr={(\thisrow{raw}-1)*100}] {breakdown.dat};
\addplot[bdLab, nodes near coords style={font=\fontsize{6}{6}\selectfont\sffamily, text=white}]
  table[y=y, x expr={(\thisrow{raw}-1)*100+min((\thisrow{ctl}-\thisrow{raw})*100, 39-(\thisrow{raw}-1)*100)/2},
  y expr={(\thisrow{ctl}-\thisrow{raw})*100>=5 ? \thisrow{y} : nan}, meta expr={(\thisrow{ctl}/\thisrow{raw}-1)*100}] {breakdown.dat};
\fill[white] (axis cs:30.4,9.62) -- (axis cs:31.6,9.62) -- (axis cs:32.2,10.38) -- (axis cs:31.0,10.38) -- cycle;
\draw[black!70, line width=0.45pt] (axis cs:30.4,9.62) -- (axis cs:31.0,10.38) (axis cs:31.6,9.62) -- (axis cs:32.2,10.38);
\node[font=\fontsize{6}{6}\selectfont\sffamily, anchor=west, inner sep=0pt] at (axis cs:41.2,10) {(HS 1.694)};
\node[font=\lpLab\bfseries, anchor=south east, inner sep=0pt] at (rel axis cs:-0.01,1.1) {(a)};
\end{axis}
\begin{axis}[bdB, xtick={0,25,50,75,100}, xlabel={Share of L2 bytes (\%)}, xbar stacked, bar width=0.44, legend columns=3, legend style={at={(0.02,1.02)}, anchor=south west, font=\fontsize{6}{7}\selectfont\sffamily, /tikz/every even column/.append style={column sep=0.9mm}}, legend image code/.code={\draw[#1] (0cm,-0.08cm) rectangle (0.24cm,0.08cm);}]
\addplot[bdLeaf] table[y expr=\thisrow{y}+0.2, x=packed] {occ-frozen.dat};
\addlegendimage{bdRawS, lpHatch}
\addplot[bdRawS, forget plot] table[y expr=\thisrow{y}+0.2, x=unpacked] {occ-frozen.dat};
\addplot[bdUp] table[y expr=\thisrow{y}+0.2, x=upper] {occ-frozen.dat};
\addplot[bdData] table[y expr=\thisrow{y}+0.2, x=data] {occ-frozen.dat};
\addplot[bdEmpty] table[y expr=\thisrow{y}+0.2, x=empty] {occ-frozen.dat};
\addlegendimage{legend image code/.code={\draw[fill=lpGrey!40, draw=lpGrey, opacity=0.7] (0cm,-0.035cm) rectangle (0.3cm,0.035cm);}}
\legend{Compressed leaves, Uncompressed leaves, Upper PT, Data, Unfilled, Baseline}
\node[font=\lpLab\bfseries, anchor=south east, inner sep=0pt] at (rel axis cs:-0.01,1.2) {(b)};
\end{axis}
\begin{axis}[bdB, hide axis, xbar stacked, bar width=0.44]
\addplot[draw=none, fill=none] table[y expr=\thisrow{y}+0.2, x=packed] {occ-frozen.dat};
\addplot[draw=none, fill=none, lpHatch] table[y expr=\thisrow{y}+0.2, x expr={\thisrow{unpacked}>=5 ? \thisrow{unpacked} : 0}] {occ-frozen.dat};
\end{axis}
\begin{axis}[bdB, hide axis, xbar stacked, bar width=0.2]
\addplot[bdLeaf, opacity=0.75] table[y expr=\thisrow{y}-0.3, x=packed] {occ-ord.dat};
\addplot[bdRawS, opacity=0.75, lpHatch] table[y expr=\thisrow{y}-0.3, x=unpacked] {occ-ord.dat};
\addplot[bdUp, opacity=0.75] table[y expr=\thisrow{y}-0.3, x=upper] {occ-ord.dat};
\addplot[bdData, opacity=0.75] table[y expr=\thisrow{y}-0.3, x=data] {occ-ord.dat};
\addplot[bdEmpty, opacity=0.75] table[y expr=\thisrow{y}-0.3, x=empty] {occ-ord.dat};
\node[font=\fontsize{6}{6}\selectfont\sffamily, text=black!60, anchor=east, inner sep=0.5pt] at (axis cs:-0.8,9.7) {base};
\end{axis}
\begin{axis}[bdB, hide axis]
\addplot[only marks, mark=none, forget plot, point meta=explicit, nodes near coords={\pgfmathprintnumber[fixed, precision=0, assume math mode=true]{\pgfplotspointmeta}\%},
  nodes near coords style={font=\fontsize{6}{6}\selectfont\sffamily, anchor=west, inner sep=0.5pt}]
  table[y expr=\thisrow{y}+0.2, x expr=100.6, meta expr={\thisrow{packed}+\thisrow{unpacked}+\thisrow{upper}}] {occ-frozen.dat};
\end{axis}
\end{tikzpicture}
\caption{Speedup breakdown and L2 occupancy on the small core, showing (a) gains from protection and from compression beyond the best static uncompressed configuration (with A, B, and C marking the selected policy) and (b) L2 occupancy under our controller (upper) versus the baseline (lower) with page-table shares labeled.}
\label{fig:breakdown}
\end{figure*}

Figure~\ref{fig:breakdown}a separates LeafPack's speedup into the gain of protecting page-table lines (captured by the best static uncompressed configuration) and the additional gain of compressing them. Protection accounts for most of the gain on standalone bc, bc+bc, and $8\times$bc, where no cache-sensitive co-runner competes for data ways, whereas compression accounts for the bulk of the gain when the active leaf set exceeds fifteen uncompressed ways (+63.7\% on 4~GiB SysBench) or when retaining raw leaves displaces co-runner data (+10.0\% on bc+omnetpp), while for gcc and omnetpp alone our controller selects the 3-way uncompressed policy C.

Figure~\ref{fig:breakdown}b shows how our controller divides L2 capacity. On bc+omnetpp, the controller caps page tables at 31\% of the L2 so application data retains 68\%, whereas on bc+bc and $8\times$bc it allows page tables to occupy 94\% of the L2 to capture the combined leaf set. Over 99\% of compressed leaf lines use our sixteen-line dense format because its 20-bit offsets span $\pm2$~GiB of physical memory. Depending on how many neighboring leaf lines a workload touches before eviction, each occupied slot holds an average of 3.0 lines on the eight-core mixes to 14.9 lines on 4~GiB SysBench, giving a median compression density of $4.7\times$ over raw storage ($2.1\times$ to $10.4\times$) in static 15-way runs. The achieved density remains below the formats' maximum compression ratios of $12.8\times$ and $6.4\times$ because occupied slots are only partially filled. In Fig.~\ref{fig:breakdown}b, \emph{Unfilled} denotes cache capacity holding no valid lines, which consists mainly of empty slots inside converted groups that application data cannot use (10.9\% of the L2 on 4~GiB SysBench, not counted in the page-table shares at right).

High compression density alone does not predict speedup, however, as standalone bc packs 6.5 lines per slot and holds 32\% of its allocated leaf footprint yet gains only 1.0\% over uncompressed protection because 15 uncompressed ways already hold its active leaf set. Instead, speedup over the best uncompressed configuration tracks the net reduction in L2 misses per 1,000 instructions (page-walk misses saved minus extra application-data misses), reaching a correlation of 0.72 without SysBench and 0.98 with SysBench included (versus 0.51 for slot occupancy alone).

\textit{LeafPack versus composing existing techniques.}
A natural alternative to LeafPack is to combine a page-table codec with an existing superblock compressed-cache layout~\cite{sardashti2016yacc} under an adaptive retention policy. Our packed-only configuration evaluates this combination by encoding each leaf line independently with LeafPack's codec in the same superblock slots under the same controller and decode latencies, so that only the cross-line format differs from full LeafPack. We also test a general-purpose per-line BDI codec~\cite{pekhimenko2012bdi} in the same layout. Across all 18 development workloads on both cores, the packed-only combination reaches 1.109 and 1.103 harmonic speedup over the baseline, sitting just 0.5\% above the best static uncompressed configuration (1.104 and 1.098) and falling below it once we exclude 4~GiB SysBench, whereas full LeafPack reaches 1.157 and 1.151 for an additional 4.3\% and 4.4\% gain. That gap comes from sharing a base descriptor across neighboring leaf lines, which no per-line codec can do and which raises packing density on our checkpoint page tables from 5.1--6.0 leaf lines per physical way under per-line packing (or 6.0--7.0 under idealized within-line packing) up to 8.3--11.3 under LeafPack. Because per-line BDI fits only about two leaf lines per slot, it trails even the best static uncompressed configuration (0.970 and 0.969) while full LeafPack leads it by 7.9\% and 8.1\%, showing that our page-table format rather than generic compression drives the speedup. Outside the development set, fragmented XSBench is the sole exception, as the two simpler formats reach 1.14 to 1.18 versus 1.05 for full LeafPack in two of the three allocator histories (Section~\ref{sec:eval-scaling}).

\textit{Remaining headroom.}
To bound how much further headroom remains, we construct an ideal leaf oracle that serves every leaf page-table read at L2-hit latency while consuming zero L2 cache capacity. Across eleven development workloads, LeafPack captures 77.8\% (small core) and 78.8\% (server core) of this ideal oracle's geometric-mean gain over the baseline, reaching 83 to 95\% on single programs and homogeneous mixes and 70 to 80\% on the omnetpp pairs, while on pr+gcc the ideal oracle gains 5.5 to 8.7\% whereas LeafPack loses about 4\% due to data-capacity contention.

\subsection{Held-Out Workloads}\label{sec:eval-heldout}

Table~\ref{tab:heldout} evaluates our frozen controller across all twenty unseen held-out mixes. LeafPack improves harmonic speedup over the baseline across all twenty mixes on both cores, reaching geometric means of 1.214 and 1.185 (and at least 1.026 on every mix) while keeping each individual co-runner at or above 0.95 of its baseline IPC on two-program pairs and 0.93 on four-program mixes. Against the best static uncompressed configuration, our controller achieves geometric-mean ratios of 1.029 and 1.021 (minimum 0.946 and 0.954 on bc+sssp), while under SRRIP data replacement the twenty mixes gain 9.9\% and 7.4\% over the SRRIP baseline and 2.8\% and 1.8\% over the best static uncompressed configuration on SRRIP.

To test generalization to applications outside our benchmark pool, the bottom of Table~\ref{tab:heldout} evaluates the DBx1000 in-memory database~\cite{yu2014dbx1000} running YCSB~\cite{cooper2010ycsb}, the Olden mst pointer-chasing benchmark~\cite{rogers1995olden}, and the SPEC lbm stencil alone and in pairs, with all workloads and pass criteria fixed before simulation. Across these six workloads on both cores, LeafPack never trails the baseline by more than 0.1\% or the best static uncompressed configuration by more than 1\%, while accelerating DBx1000 by 15.4\% and 19.6\% over the baseline (8.4\% and 10.2\% beyond uncompressed protection).

\textit{Where compression matters.} The best static uncompressed configuration is a strong reference point because picking the best fixed uncompressed ceiling offline per workload bounds any single fixed ceiling, and while it does not bound priority-based or TLB-caching protection by construction, EWaH, xPTP, and Victima at its best way cap never exceed it by more than 0.4\%, 0.3\%, and 0.1\% on any run (each under its own data replacement policy). Compression opens up a trade-off that this offline oracle cannot match, either delivering the same translation coverage in 50 to 73\% less page-table capacity (Section~\ref{sec:eval-single}) or packing more coverage into the same ways. LeafPack's lead over the oracle therefore concentrates where active leaf demand exceeds what uncompressed protection can hold, as in the held-out mixes that include 4~GiB SysBench (7.7\% and 5.7\%), or where the freed capacity returns to application data, whether for a sensitive co-runner (10.0\% on bc+omnetpp and 6.1\% on XSBench+omnetpp) or for the program itself (8.4\% and 10.2\% on standalone DBx1000), while on the remaining held-out mixes it matches the oracle on average (+0.4\% and +0.2\%).

\begin{table}[b]
\caption{\textbf{Held-out mixes and new benchmarks} (\emph{HS} / \emph{vs.\ best static uncompressed}).}
\label{tab:heldout}
\centering
\footnotesize
\setlength{\tabcolsep}{4pt}
\begin{tabular}{@{}lcc@{}}
\toprule
Workload & Small core & Server core \\
 & \multicolumn{2}{c}{\scriptsize (HS / vs.\ best static uncompressed)} \\
\midrule
bfs+XSBench                  & 1.151 / 1.007 & 1.163 / 1.012 \\
bfs+bfs                      & 1.286 / 1.024 & 1.250 / 1.027 \\
XSBench+XSBench              & 1.317 / 1.021 & 1.261 / 0.985 \\
pr+sssp                      & 1.204 / 0.960 & 1.184 / 0.957 \\
bc+bfs                       & 1.350 / 1.049 & 1.303 / 1.045 \\
bc+sssp                      & 1.175 / 0.946 & 1.146 / 0.954 \\
bfs+SysBench 4\,GiB          & 1.038 / 1.017 & 1.038 / 1.015 \\
bc+SysBench 4\,GiB           & 1.416 / 1.213 & 1.386 / 1.193 \\
pr+SysBench 4\,GiB           & 1.335 / 1.174 & 1.273 / 1.109 \\
sssp+bfs                     & 1.067 / 0.988 & 1.049 / 0.997 \\
pr+pr                        & 1.196 / 1.002 & 1.202 / 1.009 \\
bc+XSBench                   & 1.352 / 1.045 & 1.383 / 1.088 \\
sssp+XSBench                 & 1.310 / 1.001 & 1.269 / 0.987 \\
pr+XSBench                   & 1.280 / 1.001 & 1.225 / 1.003 \\
\midrule
pr+sssp+XSBench+SysBench     & 1.280 / 1.062 & 1.218 / 1.034 \\
bc+bfs+SysBench+omnetpp      & 1.069 / 1.028 & 1.044 / 1.012 \\
pr+sssp+bfs+XSBench          & 1.235 / 0.986 & 1.207 / 0.988 \\
pr+XSBench+SysBench+omnetpp  & 1.159 / 1.038 & 1.105 / 1.026 \\
pr+bfs+SysBench+omnetpp      & 1.064 / 1.023 & 1.026 / 1.019 \\
sssp+bfs+XSBench+omnetpp     & 1.099 / 1.022 & 1.061 / 0.982 \\
\midrule
Geometric mean (held-out) & 1.214 / 1.029 & 1.185 / 1.021 \\
\midrule
DBx1000     & 1.154 / 1.084 & 1.196 / 1.102 \\
mst         & 1.000 / 1.000 & 1.000 / 1.000 \\
lbm         & 1.001 / 1.001 & 1.000 / 1.000 \\
DBx1000+lbm & 1.047 / 1.010 & 1.050 / 1.010 \\
mst+lbm     & 1.006 / 0.999 & 1.005 / 1.000 \\
DBx1000+mst & 0.999 / 0.990 & 1.023 / 0.991 \\
\midrule
Geometric mean (new) & 1.033 / 1.014 & 1.044 / 1.016 \\
\bottomrule
\end{tabular}
\end{table}

\subsection{Replacement and Translation Baselines}\label{sec:eval-baselines}

\textit{SRRIP and Victima.}
To test a stronger replacement policy and compare fairly with Victima~\cite{victima}, Fig.~\ref{fig:dev}b repeats our nine headline workloads using 3-bit SRRIP for application data. Because the SRRIP baseline improves performance by 1.8 to 21.9\% over random replacement, LeafPack's gains over our default random-replacement baseline exceed its gains over the SRRIP baseline on these workloads, so we report both. Even on top of SRRIP, LeafPack delivers further gains, improving geometric-mean harmonic speedup over SRRIP by 18.7\% on the small core and 17.7\% on the server core across the nine headline workloads (and by 9.5\% and 8.7\% over the SRRIP baseline, or 4.5\% and 4.4\% over the best static uncompressed configuration on SRRIP, across all 18 development workloads).

When we evaluate Victima on SRRIP using its best L2 way cap per workload, it achieves 0.90 to 1.03 of the SRRIP baseline on the headline workloads, whereas LeafPack on SRRIP outperforms Victima on every headline workload on both cores. Across thirteen 1~MiB workloads, Victima reaches 0.90 to 1.06 of the SRRIP baseline while LeafPack leads Victima by 13.7\% and 14.9\% in geometric mean. We also reduced our default page-walker count from 32 to 4 to increase the cost of page-table walks, which left this comparison unchanged. With a 2~MiB L2 cache, Victima breaks even on bc (1.004), while LeafPack still leads Victima by 20\% and 21\% across five workloads. This difference stems from walk cost in our Arm hierarchy, where 92\% of L2 TLB misses hit a partial walk-cache entry in the L2 TLB and require only a single leaf read from the L2 cache. When we instead configure three 32-entry dedicated walk caches without L2-TLB partial translations to match Victima's original setup, Victima gains 6.1\% on XSBench as expected, while LeafPack leads Victima by 4 to 6\% on XSBench and 15 to 17\% on bc.

\textit{Prioritization and equal-area alternatives.}
We model EWaH and xPTP~\cite{ewah,xptp} as L2 replacement policies and compare each design against the baseline with the same data replacement policy, since xPTP relies on LRU recency order. While protecting page-table lines under EWaH speeds up standalone bc (1.27 HS), it hurts most co-runs with data-sensitive applications, dropping performance to 0.48 to 0.99 of the baseline under random replacement, 0.34 to 0.98 under SRRIP, and 0.26 to 0.99 under LRU. By turning off page-table priority when translation pressure is low, xPTP avoids most of that slowdown (0.83 at worst) but also gives up most of the translation speedup, reaching only 1.01 on 4~GiB SysBench and 1.17 on bc. Even when we compare against a per-workload oracle that picks the best prioritization scheme or the baseline for each run, LeafPack leads in geometric mean over 17 development workloads by 9.4\% and 9.3\% under random replacement, 6.4\% and 6.1\% under SRRIP (over EWaH or the baseline), and 5.6\% and 4.9\% under LRU (over EWaH, xPTP, or the baseline), falling below the LRU baseline only on the two graph and gcc co-runs (0.85 to 0.98). We also compare LeafPack against spending a comparable or larger area budget (at least 19.4~KiB or 0.050~mm$^2$, within 14 to 21\% of our design's 0.058 to 0.063~mm$^2$) on conventional structures across bc, XSBench, SysBench, and bc+omnetpp. Doubling or quadrupling the L2 TLB (to 2,560 or 5,120 entries per core with one extra access cycle), adding a 16~KiB dedicated walk cache (3.1 to 3.4$\times$ our area), or adding an extra L2 way (1.2 to 1.3$\times$ our area) yields only 0.997 to 1.034 harmonic speedup, compared to 1.18 to 1.69 for LeafPack.

\textit{CoLT.}
Because CoLT~\cite{colt} coalesces TLB entries only when consecutive virtual pages map strictly contiguous physical frames, which occurs infrequently in our interleaved multiprogrammed boots, CoLT alone achieves 0.995 to 1.058 harmonic speedup. However, combining LeafPack with CoLT preserves LeafPack's speedup within 0.7\% (never more than 0.3\% below it) across contiguous, interleaved, and fragmented mappings and never trails CoLT alone.

\subsection{Scaling and Sensitivity}\label{sec:eval-scaling}

\begin{figure}[b]
\centering
\pgfplotsset{snbar/.style={lpbase, scale only axis, clip=false, height=1.3cm, ymin=0, ymax=74, ytick={0,20,40,60},
  ybar, bar width=0.4, bar shift=0pt, xmajorgrids=false, ymajorgrids=true, axis x line*=bottom, axis y line*=left,
  every axis plot/.append style={draw=white, line width=0.25pt}, xtick=\empty, xtick style={draw=none}},
  snA/.style={snbar, width=3.5cm, xmin=-1.6, xmax=13.6}, snB/.style={snbar, width=2.8cm, xmin=-1.3, xmax=10.3, bar width=0.5},
  snNamesA/.style={snNames, xtick={0,4,8,12}}, snNamesB/.style={snNames, xtick={0,3,6,9}, xticklabels={bc, {XS-\\Bench}, {Sys-\\Bench}, {bc+\\omnet.}}},
  snNames/.style={xticklabels={bc, XSBench, SysBench, {bc+\\omnet.}}, x tick label style={font=\fontsize{6}{7}\selectfont\sffamily, align=center, anchor=north, inner sep=1.5pt}}}
\tikzset{snRaw/.style={fill=lpVerm!45, draw=lpVerm!80!black, line width=0.3pt, lpHatch}, snCtl/.style={fill=black},
  snRawN/.style={fill=lpVerm!10, draw=lpVerm!70, line width=0.3pt, postaction={pattern={Lines[angle=45, distance=2.8pt, line width=0.45pt]}, pattern color=lpVerm!55}}, snCtlN/.style={fill=black!30},
  snR1/.style={fill=lpVerm!12, draw=lpVerm!70, line width=0.3pt, postaction={pattern={Lines[angle=45, distance=2.8pt, line width=0.45pt]}, pattern color=lpVerm!60}},
  snR2/.style={fill=lpVerm!45, draw=lpVerm!80!black, line width=0.3pt, lpHatch},
  snR4/.style={fill=lpVerm!85!black, draw=lpVerm!60!black, line width=0.3pt, postaction={pattern={Lines[angle=45, distance=2.8pt, line width=0.45pt]}, pattern color=white}},
  snC1/.style={fill=black!28}, snC2/.style={fill=black!58}, snC4/.style={fill=black}}
\newcommand{\snAplots}[3]{
\addplot[snR1] table[x expr={#2-1-0.21}, y expr={\thisrow{mib}==1 ? (\thisrow{raw#1}-1)*100 : nan}] {cap-#3.dat};
\addplot[snR2] table[x expr={#2-0.21},   y expr={\thisrow{mib}==2 ? (\thisrow{raw#1}-1)*100 : nan}] {cap-#3.dat};
\addplot[snR4] table[x expr={#2+1-0.21}, y expr={\thisrow{mib}==4 ? (\thisrow{raw#1}-1)*100 : nan}] {cap-#3.dat};
\addplot[snC1] table[x expr={#2-1+0.21}, y expr={\thisrow{mib}==1 ? (\thisrow{ctl#1}-1)*100 : nan}] {cap-#3.dat};
\addplot[snC2] table[x expr={#2+0.21},   y expr={\thisrow{mib}==2 ? (\thisrow{ctl#1}-1)*100 : nan}] {cap-#3.dat};
\addplot[snC4] table[x expr={#2+1+0.21}, y expr={\thisrow{mib}==4 ? (\thisrow{ctl#1}-1)*100 : nan}] {cap-#3.dat};}
\newcommand{\snA}[2]{%
\begin{axis}[snA, #2]
\snAplots{#1}{0}{bc}\snAplots{#1}{4}{XSBench}\snAplots{#1}{8}{SysBench}\snAplots{#1}{12}{bc-omnetpp}
\end{axis}}
\newcommand{\snB}[2]{%
\begin{axis}[snB, #2]
\addplot[snRawN] table[x expr={3*(\thisrow{x}-1)-0.81}, y expr={(\thisrow{natRaw#1}-1)*100}] {nested.dat};
\addplot[snCtlN] table[x expr={3*(\thisrow{x}-1)-0.27}, y expr={(\thisrow{natCtl#1}-1)*100}] {nested.dat};
\addplot[snRaw] table[x expr={3*(\thisrow{x}-1)+0.27}, y expr={(\thisrow{nstRaw#1}-1)*100}] {nested.dat};
\addplot[snCtl] table[x expr={3*(\thisrow{x}-1)+0.81}, y expr={(\thisrow{nstCtl#1}-1)*100}] {nested.dat};
\end{axis}}
\begin{tikzpicture}
\snA{S}{}
\node[font=\lpLab\bfseries, anchor=south west, inner sep=0pt] at (0,1.42cm) {(a) L2 size};
\snA{V}{at={(0,-1.55cm)}, snNamesA}
\node[font=\lpLab, rotate=90, anchor=south] at (-1.05cm,-0.12cm) {Speedup (\%)};
\node[font=\lpLeg\bfseries, rotate=90, anchor=south] at (-0.62cm,0.65cm) {Small core};
\node[font=\lpLeg\bfseries, rotate=90, anchor=south] at (-0.62cm,-0.9cm) {Server core};
\begin{scope}[xshift=4.05cm]
\snB{S}{}
\node[font=\lpLab\bfseries, anchor=south west, inner sep=0pt] at (0,1.42cm) {(b) Nested paging};
\snB{V}{at={(0,-1.55cm)}, snNamesB}
\end{scope}
\end{tikzpicture}
\\[0.3mm]{\lpLeg\begin{tabular}{@{}l@{\hspace{1.5mm}}l@{\hspace{3mm}}l@{}}
(a) 1/2/4 MiB, light to dark: & \tikz{\draw[snR1] (0,0) rectangle (0.2,0.18); \draw[snR2] (0.22,0) rectangle (0.42,0.18); \draw[snR4] (0.44,0) rectangle (0.64,0.18);}~Best static uncompr. & \tikz{\fill[snC1] (0,0) rectangle (0.2,0.18); \fill[snC2] (0.22,0) rectangle (0.42,0.18); \fill[snC4] (0.44,0) rectangle (0.64,0.18);}~LeafPack \\
(b) native: & \tikz\draw[snRawN] (0,0) rectangle (0.3,0.18);~Best static uncompr. & \tikz\fill[snCtlN] (0,0) rectangle (0.3,0.18);~LeafPack \\
(b) nested: & \tikz\draw[snRaw] (0,0) rectangle (0.3,0.18);~Best static uncompr. & \tikz\fill[snCtl] (0,0) rectangle (0.3,0.18);~LeafPack
\end{tabular}}
\caption{Sensitivity to L2 capacity and nested paging, showing harmonic speedup gain (\%) over a same-size baseline across (a) 1, 2, and 4~MiB L2 caches (light to dark) and (b) native (light) versus nested (dark) paging.}
\label{fig:sens}
\end{figure}
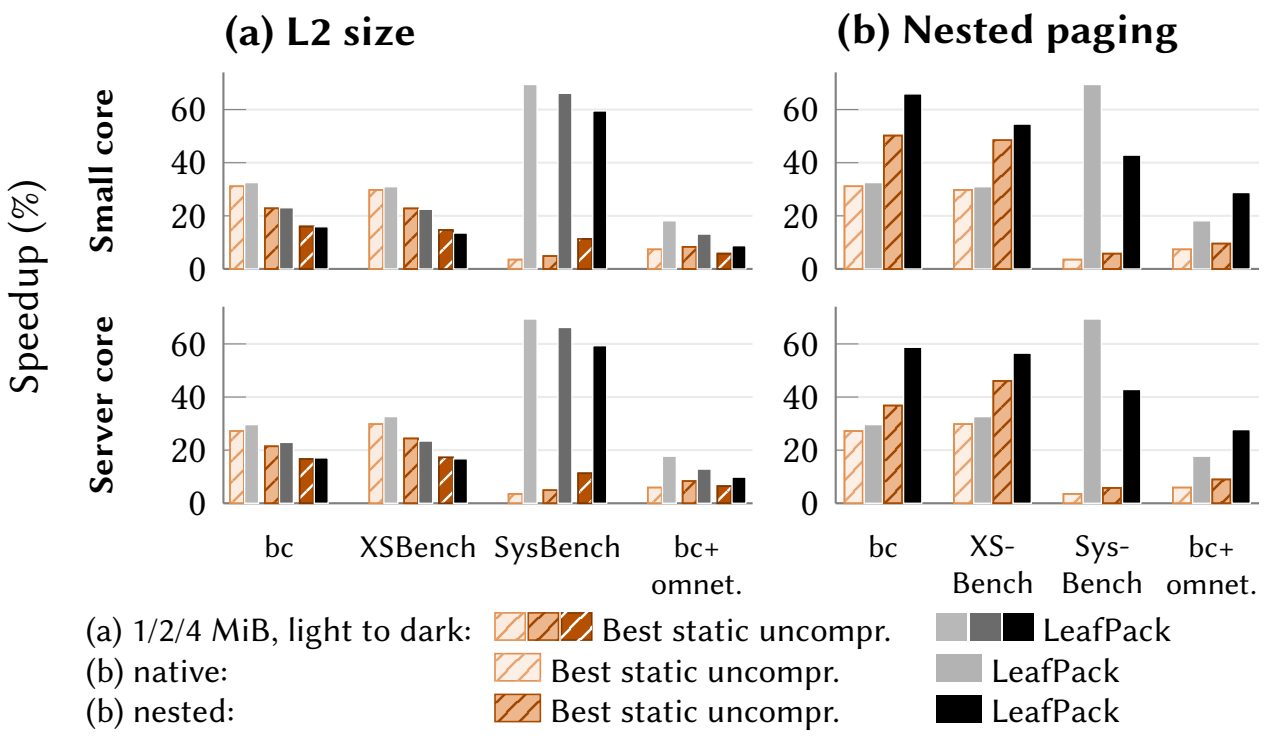

\textit{Core count and cache size.}
Scaling from one copy of bc on a 1~MiB L2 to eight copies sharing a 4~MiB L2 multiplies leaf demand by eight while only quadrupling cache capacity. At a 5-way ceiling, compression cuts leaf L2 misses from 34.1 to 11.2 per 1,000 instructions (37.2 to 12.3 on the server core) and gains 15.2\% and 14.9\% over 5-way uncompressed protection, while our controller selects 15 compressed ways to reach 1.331 and 1.321 harmonic speedup. Conversely, enlarging the L2 from 1 to 4~MiB for a fixed workload allows more leaves to fit uncompressed (Fig.~\ref{fig:sens}a), yet LeafPack never falls more than 1\% below the best static uncompressed configuration on bc and XSBench at 2 and 4~MiB while still leading it by 2.8 to 4.4\% on bc+omnetpp and by 43 to 58\% on 4~GiB SysBench.

\textit{Physical contiguity and fragmentation.}
Because LeafPack relies on physical frame regularity, booting bc+bc with contiguous rather than interleaved page allocation raises 5-way compressed speedup from 1.264/1.230 to 1.363/1.317, whereas eight-core bc sees no gain from initial contiguity because each process repeatedly frees and reallocates two 64~MiB arrays per iteration, interleaving frames at run time. Under heavier fragmentation across three allocator histories generated by a synthetic fragmenter, our frozen controller stays within 0.7\% of the best static setting on fragmented bc and 4~GiB SysBench. Fragmented XSBench exposes a limitation, as the controller selects its 3-way uncompressed candidate in two histories to reach 1.04 to 1.05 harmonic speedup on both cores, or 0.82 of the best static setting (1.28 at 10 ways, outside the candidate set), and reaches 0.955 and 0.926 of it in the third history. Because our evaluated controller charges every leaf-line miss to the leader set of the line's packed index even when an incompressible line lives in its normal set, 99.6\% and 99.1\% of the counted incompressible misses on fragmented XSBench land on compressed leader sets and push the controller toward the 3-way uncompressed policy. When we evaluate an ablation that instead charges each miss to the set that stores the line and delays compressed probing until four times the warm-up period (adding two counter bits), fragmented XSBench rises to 1.21 and 1.18 across all three histories while 21 other development and fragmented runs stay within 1\% of our evaluated controller. We audited this rule across 108,448 counted misses, including directed tests that force policy flips, probe-mode transitions, and in-flight miss merges, and confirmed zero misattributed misses. Because we designed this change after our held-out evaluation, we report it here as an ablation and keep the evaluated controller unchanged.

\textit{Three-level hierarchy.}
Adding a private 256~KiB, 8-way L2 to each core while giving the shared L3 our default L2 geometry halves the leaf reads that reach the L3. Across the nine headline workloads and two gcc pairs in this hierarchy, LeafPack improves geometric-mean performance by 24.3\% and 22.9\% over the baseline and by 7.0\% and 7.1\% over the best static uncompressed configuration, closely matching their two-level results (7.5\% and 7.6\%). Because the private L2s also absorb most of gcc's data pressure, those two gcc pairs gain 0.4 to 1.5\% here instead of losing about 4\% as in the two-level system.

\textit{Huge pages, nesting, and translation hierarchy.}
When we sweep the L2 TLB from 640 to 5,120 entries and either disable the L2 prefetcher or replace it with a virtual-address stride and indirect prefetcher, LeafPack's speedup and its margin over uncompressed protection remain essentially unchanged across bc, XSBench, SysBench, and bc+omnetpp. When transparent huge pages~\cite{linux_transhuge} cover 98.9 to 99.9\% of mapped memory, the allocated 4~KiB leaf footprint shrinks to 8 to 95~KiB and LeafPack stays within 0.6\% of uncompressed protection (1.000 to 1.036 HS). Conversely, under two-dimensional nested paging with 4~KiB guest and host pages where LeafPack compresses leaf lines at both stages (using a fixed stage-2 identity map without VM exits), our controller improves harmonic speedup by 28 to 66\% over the baseline and by 4 to 35\% over the best static uncompressed configuration (Fig.~\ref{fig:sens}b).

\textit{Latency, port contention, and seeds.}
LeafPack's gains are also insensitive to L2 datapath timing and simulation methodology. Eliminating our two- and three-cycle decode latency changes IPC by only about 0.2\%, while adding an extra decode cycle or a two-cycle fill encoder changes harmonic speedup by at most 0.24\%. Modeling a single-ported L2 array that serializes slot metadata and payload reads, converted-group probes, and dirty reclamation writebacks (without charging for clean victims or tag updates) changes harmonic speedup by only $-0.7$ to $+0.4\%$ across eight workloads. Locking the set during group transitions and capping transition writebacks at two write-buffer entries similarly changes harmonic speedup by at most 0.2\% across 11 workloads on both cores. Evaluating workloads at fixed instruction counts rather than fixed cycle windows preserves all policy rankings across six tested workloads (changing LeafPack's margin over uncompressed protection by at most 3\%), and varying the random seed across four leader-set hashes and replacement generators (eight for sssp+omnetpp and pr+gcc) changes harmonic speedup by only 0.1 to 0.8\% on bc, SysBench, and Mix-3 and by 1.6\% on pr+gcc. The sole exception is sssp+omnetpp, where one of the eight leader-set hashes causes the controller to select the 3-way uncompressed policy throughout the run and settle at 1.041 and 1.028 instead of 1.111 and 1.082.

\subsection{Limitations and Policy Trade-offs}\label{sec:discussion}

\textit{Co-runs with gcc.}
Across our development set, sssp+gcc and pr+gcc are the only workloads that trail the baseline by more than 0.5\%. On both pairs, our controller selects the 5-way compressed policy and reaches 0.957 to 0.962 HS, trailing the 3-way uncompressed policy (0.988 to 1.004 HS). On pr+gcc, compressing leaf lines eliminates nearly all of pr's walk misses (1.37M to 0.01M), allowing pr to execute faster and issue 19\% more data misses while gcc makes 20\% less progress (1.16M versus 1.45M instruction accesses). This slowdown stems from L2 data capacity contention rather than memory bandwidth, as dedicating even three uncompressed ways to page tables reduces gcc to 0.91 to 0.96 of its progress under the baseline, and expanding to five compressed ways raises gcc's L2 misses per instruction by another 15 to 25\% while DRAM read latency and bus utilization change by at most 6\% and 3.3 percentage points. Our controller picks the 5-way policy on these two pairs because its shadow tags observe the access stream generated under the compressed leader sets rather than a physical uncompressed cache, which overestimates the 3-way policy's data misses by 6 to 8\% and reverses a true difference of 0 to 5\%. On every other development workload and core, our controller selects the uncompressed policy whenever it lies within 1\% of the best.

\textit{Choice of candidate policies.}
Our two compressed candidates serve two distinct operating regimes. The 15-way ceiling excels on large active leaf sets, reaching 1.71 versus 1.11 static harmonic speedup on 4~GiB SysBench and 1.38 versus 1.26 on bc+bc while occupying 66 to 94\% of the L2. Conversely, the 5-way ceiling caps page-table occupancy at 31\% of the L2 to protect data-sensitive co-runners, reaching 1.18 versus 1.02 on bc+omnetpp and 0.97 versus 0.46 on sssp+gcc. While a 4-way partial group improves Mix-2 and the gcc pairs slightly, we restrict our controller to three candidate policies because evaluating a fourth candidate would require dedicating another 32 physical leader sets and adding more pairwise comparison counters.

\textit{Scope and future work.}
LeafPack targets the page walks that remain after huge pages, where memory stays PTE-mapped yet retains short runs of consecutive physical frames. Because each compressible leaf line needs only eight consecutive frames (32~KiB) rather than an aligned 2~MiB block, this regime persists when operators disable transparent huge pages for tail latency~\cite{redis_latency}, when the kernel splits a huge-page mapping without breaking up its backing allocation, and under modern multi-size huge pages that remain PTE-mapped~\cite{linux_thp_admin,linux_thp_mm}. Our native Arm experiments with 4~KiB pages evaluate this regime directly. Two-dimensional page walks expand the active leaf working set even further, and our nested-paging proof of concept in Section~\ref{sec:eval-scaling} improves harmonic speedup over the baseline by 28 to 66\% under a fixed stage-2 identity map without VM exits. We leave modern-kernel huge-page policies, full virtualization, and a held-out evaluation of our fill-set attribution ablation to future work.

\section{Related Work}\label{sec:related}

\textit{Page-table and translation caching.}
Page-walk caches keep upper-level descriptors close to the MMU~\cite{barr2010,bhattacharjee2013} while leaving leaf descriptors to the data caches. Like EWaH and xPTP~\cite{ewah,xptp} (Section~\ref{sec:background}), LeafPack retains leaf lines in the shared cache, but compresses regular leaf lines into shared slots under a bounded way ceiling to preserve data capacity. By contrast, rather than caching page-table lines, Victima~\cite{victima} stores completed translations in L2 blocks, DUCATI~\cite{ducati} lets a GPU last-level cache hold TLB entries alongside data and backs them with a TLB in DRAM, POM-TLB~\cite{ryoo2017pomtlb} places a large translation table in DRAM whose entries are cached in the data hierarchy, and near-memory walkers execute walks inside the memory controller~\cite{picorel2017dipta}. By compressing and caching the architectural page-table lines themselves instead of maintaining a secondary store of completed translations, LeafPack leaves the walk algorithm and TLB shootdowns unchanged.

\textit{TLB reach, contiguity, and prefetching.}
Coalescing and clustered TLBs combine contiguous or nearby page translations into single TLB entries~\cite{colt,pham2014clustered,park2017hybrid,talluri1994subblocking}, and operating-system and memory-allocator policies increase physical contiguity to expose more coalescing opportunities~\cite{alverti2020spot,margaritov2021ptemagnet,navarro2002superpages,kwon2016ingens,panwar2018illuminator,panwar2019hawkeye,zhu2020quicksilver,temeraire}. Whenever a workload's active footprint exceeds the reach of the coalesced TLBs, however, TLB misses still trigger hardware page-table walks through the cache hierarchy. Translation prefetchers that predict upcoming translations~\cite{vavouliotis2021agile,vavouliotis2021morrigan,margaritov2019asap,bhattacharjee2017tempo,bhattacharjee2010icc,kandiraju2002distance} and speculative address translation designs~\cite{revelator} similarly fetch their page table descriptors through the data caches. Because Trail~\cite{kanellopoulos2026trail} also embeds per page prefetch deltas in unused PTE bits, it can prevent LeafPack from compressing a leaf line whenever those bits differ across its eight PTEs. We exploit this spatial regularity inside the shared cache so that both demand and prefetched walks hit on chip, composing cleanly with CoLT (Section~\ref{sec:eval-baselines}) and benefiting from contiguous operating-system page allocation (Section~\ref{sec:eval-scaling}).

\textit{Alternative page-table organizations.}
Elastic cuckoo page tables replace dependent radix tree lookups with parallel hash lookups~\cite{skarlatos2020}, Compendia selectively merges radix-tree levels~\cite{compendia}, and Utopia and direct segment-like translation schemes constrain virtual-to-physical placement so large regions can be translated in a single step~\cite{kanellopoulos2023utopia,alam2017dvmt,zhang2024dmt,basu2013directsegments,karakostas2015rmm}. Virtualized architectures similarly accelerate two-dimensional guest-to-host walks through nested walk caches or flattened intermediate tables~\cite{bhargava2008,ahn2012,gandhi2014,gandhi2016agile}. Whereas those designs modify both the hardware page-table walkers and the operating system's memory manager, LeafPack keeps the standard radix page-table layout and operating-system interface unchanged.

\textit{Cache compression and adaptive management.}
General-purpose compression algorithms such as BDI, FPC, C-Pack, FVC, SC$^2$, and HyComp encode arbitrary data blocks~\cite{pekhimenko2012bdi,alameldeen2004fpc,chen2010cpack,yang2000fvc,arelakis2014sc2,arelakis2015hycomp}, while compressed-cache layouts such as DCC and SCC track neighboring compressed blocks under shared superblock tags~\cite{sardashti2013dcc,sardashti2014scc}, and YACC~\cite{sardashti2016yacc} adds per-block coherence state. Rather than expanding page table residency in the caches, TMCC~\cite{panwar2022tmcc} and DyLeCT~\cite{dylect} accelerate the physical to DRAM address lookup required by hardware compressed main memory, with TMCC storing compression metadata inside page table blocks. Building on superblock tag sharing, we target shared-cache page-table residency by managing compressed leaf lines as a dedicated cache class under an adaptive way ceiling and tailoring our packed ($6.4\times$) and dense ($12.8\times$) formats to regular page-table descriptors within five-way groups. Adaptive compressed caches decide at runtime whether compression pays off~\cite{alameldeen2004fpc}, while CAMP uses compressed block size to guide replacement~\cite{camp}. Our runtime controller builds on set dueling~\cite{10.1145/1250662.1250709,jaleel2008tadip} and shared-cache partitioning~\cite{qureshi2006ucp,suh2004partitioning,changccp,xie2009pipp,sanchez2011vantage} by adding shadow-tag evaluation and per-core miss weighting.

\section{Conclusion}\label{sec:conclusion}

We present LeafPack, a shared-cache design that packs regular leaf page-table lines into compact groups while preserving architectural page-table formats, per-line coherence, and uncompressed fallback for irregular lines. By combining our cross-line descriptor format with an online controller that adapts the page-table way ceiling, LeafPack either retains larger leaf working sets in the cache or delivers the same translation coverage in fewer ways so co-running applications keep more room for their data. Across our nine translation-heavy workloads on small and server cores, LeafPack improves geometric-mean harmonic speedup over the baseline by 28.7\% and 28.4\% and speeds up all twenty held-out mixes.

\noindent\textbf{Acknowledgments.} We used generative AI tools (OpenAI ChatGPT and Codex, and Anthropic Claude) to assist with drafting and revising text, code, experimental scripts, tables, and figures. We reviewed all generated material, verified all sources and results, and take full responsibility for the work.

\clearpage
\bibliographystyle{ACM-Reference-Format}
\bibliography{refs}

\end{document}